\documentclass[fleqn,usenatbib]{mnras}

\usepackage[T1]{fontenc}
\usepackage{ae,aecompl}

\usepackage{eso-pic}

\AtBeginShipout{%
  \ifnum\value{page}>1 %
    \typeout{* Additional boxing of page `\thepage'}%
    \setbox\AtBeginShipoutBox=\hbox{\copy\AtBeginShipoutBox}%
  \fi
}

\usepackage{graphicx}	
\usepackage{amsmath}	
\usepackage{amssymb}	
\usepackage[shortlabels]{enumitem}

\usepackage{multicol}

\title[MaNGA: Properties of S0 galaxies]{Galaxy properties as revealed by MaNGA. III. Kinematic profiles and stellar population gradients in S0s}

\author[ Dom\'inguez S\'anchez et al.]{
\parbox{\textwidth}{
  H.~Dom\'{i}nguez S{\'a}nchez$^{1,2}$\thanks{E-mail: \texttt{\rm \texttt{dominguez@ice.csic.es}}}, M.~Bernardi$^{1}$\thanks{\texttt{\rm \texttt{bernardm@sas.upenn.edu}}}, F. Nikakhtar$^{1}$, B. Margalef-Bentabol$^{1}$ and R.~K.~Sheth$^{1}$}\\
 \vspace{0.cm}\\~\\
$^{1}$ Department of Physics and Astronomy, University of Pennsylvania, Philadelphia, PA 19104, USA\\
$^{2}$ Institute of Space Sciences (ICE, CSIC), Campus UAB, Carrer de Magrans, E-08193 Barcelona, Spain\\
}

\pubyear{2020}

\begin{document}
\label{firstpage}
\pagerange{\pageref{firstpage}--\pageref{lastpage}}
\maketitle

\begin{abstract}
  This is the third paper of a series where we study the stellar population gradients (SP; ages, metallicities, $\alpha$-element abundance ratios and stellar initial mass functions)  of early type galaxies (ETGs) at $z\le 0.08$ from the MaNGA-DR15 survey. In this work we focus on the S0 population and quantify how the SP varies across the population as well as with galactocentric distance. We do this by measuring Lick indices and comparing them to stellar population synthesis models.  This requires spectra with high signal-to-noise which we achieve by stacking in bins of luminosity (L$_r$) and central velocity dispersion ($\sigma_0$). We find that: 1) There is a bimodality in the S0 population: S0s more massive than $3\times 10^{10}M_\odot$ show stronger velocity dispersion and age gradients (age and $\sigma_r$ decrease outwards) but little or no metallicity gradient, while the less massive ones present relatively flat age and velocity dispersion profiles, but a significant metallicity gradient (i.e. [M/H] decreases outwards). Above $2\times10^{11}M_\odot$ the number of S0s drops sharply. These two mass scales are also where global scaling relations of ETGs change slope. 2) S0s have steeper velocity dispersion profiles than fast rotating elliptical galaxies (E-FRs) of the same luminosity and velocity dispersion.  The kinematic profiles and stellar population gradients of E-FRs are both more similar to those of slow rotating ellipticals (E-SRs) than to S0s, suggesting that E-FRs are not simply S0s viewed face-on. 3) At fixed $\sigma_0$, more luminous S0s and E-FRs are younger, more metal rich and less $\alpha$-enhanced. Evidently for these galaxies, the usual statement that `massive galaxies are older' is not true if $\sigma_0$ is held fixed.
  
\end{abstract}




\section{Introduction}\label{sec:intro}
%

The properties of early type galaxies (hereafter ETGs) provide insight into their assembly and evolutionary paths, being  crucial for our understanding of galaxy formation. In particular, the formation mechanisms of the S0 population are still debated. Many key properties of the stellar population of a galaxy -- the total mass in stars, the mean stellar age, star formation history, chemical composition (metallicity, $\alpha$-element abundance ratios, dust content) and the Initial Mass Function (hereafter IMF) -- are encoded in its spectrum \citep{Worthey1994, Trager1998, Kauffmann2003, Bernardi2006, Panter2007, TMJ2011, Vazdekis2015, TW2017, Conroy2018}.

While there are clear correlations between the stellar populations of galaxies and their morphology, even amongst just ETGs, the stars in more massive galaxies tend to be older, more metal rich, more $\alpha$-enhanced (implying that they formed their stars over a shorter timescale) \cite[e.g.][]{Thomas2005}, with velocity dispersion driving many of the observed correlations \cite[e.g.][]{Bernardi2005}.  These changes contribute to the observed scaling relations between half-light radius, velocity dispersion and stellar mass.  In particular, curvature in these scaling relations \cite[e.g.][]{Bernardi2011} is thought to indicate changes in the stellar population and assembly history across the early-type population.  


Moreover, there are stellar population variations even within a single galaxy. The  stellar population gradient (i.e., the correlation between stellar population and distance from the center) contains hints about the assembly history and is expected to constrain inside-out versus outside-in and in-situ versus ex-situ formation scenarios \citep{Parikh2019,Ferreras2019,Lacerna2020}. These gradients are revealed thanks to Integral Field Unit surveys -- such as the one used in this work, MaNGA (\citealt{Bundy2015}) -- which provide spatially resolved spectra for individual galaxies. 

One of the goals of the present study is to see if we can decode the story that is encoded both in the curvature of scaling relations and in SP gradients.  In Papers~I and~II of this series \citep{DS2019, BDS2019} we analyzed the stellar population gradients from stacked spectra of slow and fast rotator elliptical galaxies (hereafter E-SRs and E-FRs).  The main goal here is to study the gradients in S0s.

Previous work suggests that whereas more luminous S0s have been formed at high redshift by violent, dissipative processes, less luminous lenticulars have been formed by secular processes gently stripping gas from spiral discs \citep{Barway2007, Barway2009}.
More recent work suggests that the bulges of high-mass, old, and metal-rich S0s are older than their discs, while the bulges of young, metal-poor and lower mass S0s are slightly younger than their discs \citep[FM2018 hereafter]{mangaS0}.  We would like to see how our data align with these findings.  Our study is also motivated by the suggestion that the distinction between fast rotator ellipticals and S0s is artificial:  that E-FRs are simply S0s viewed face-on \citep{C16, Graham2019}.  The stacked spectra in Papers~I and~II did not allow a clear test of this assertion.  Therefore, a secondary goal here is to check if the kinematic profiles, stellar populations and SP gradients of E-FRs and S0s are similar.

Section~\ref{sec:data} discusses how we select our sample, and the relevant photometric, spectroscopic and morphological information available to us.  It also compares the global scaling relations (how size and velocity dispersion vary with stellar mass) of S0s with those of E-FRs and E-SRs.  Whereas this comparison does not use any information from gradients or stellar population modeling, Section~\ref{sec:grads} measures gradients in the kinematics and the chemical abundances of S0s.  It then uses the measured chemical abundances to constrain the stellar populations (light-weighted age, metallicity, $\alpha$-enhancement, IMF) of S0s.  To check if E-FRs are simply face-on S0s, Section~\ref{sec:E+S0s} performs a similar analysis of E-FRs which are selected to have similar luminosities and velocity dispersions as our S0s.  Section~\ref{sec:previous} discusses our results in the context of previous work, especially how our analysis of gradients compares with the bulge-disk analysis of FM2018.  A final section summarizes our findings.  

\section{Data}\label{sec:data}

\subsection{MaNGA survey and photometry}
The MaNGA survey (\citealt{Bundy2015,Drory2015,Law2015,Law2016,Yan2016a, Yan2016b}) is a component of the Sloan Digital Sky Survey IV (\citealt{Gunn2006, Smee2013, Blanton2017}; hereafter SDSS IV) using integral field units (IFUs) to measure spectra across $\sim$ 10000 nearby galaxies. The MaNGA selection function, while complicated, is well defined \citep{Wake2017}. In what follows, when we refer to `weighted' quantities, the weight is {\tt ESWEIGHT} from \cite{Wake2017}. 

In this work, as in Papers~I and~II, we use the MaNGA DR15  \citep{Westfall2019,Aguado2019,Fischer2019}. The photometric parameters we use come from the PyMorph Photometric Value Added Catalogue (MPP-VAC) presented in \citet{Fischer2019}, which provides photometric parameters from S\'ersic and S\'ersic + Exponential fits (we use the `truncated' magnitudes and sizes). We refer the reader to the corresponding references for further details on the observations and photometric reductions.

\begin{figure*}
  \centering
  \includegraphics[width=0.8\linewidth]{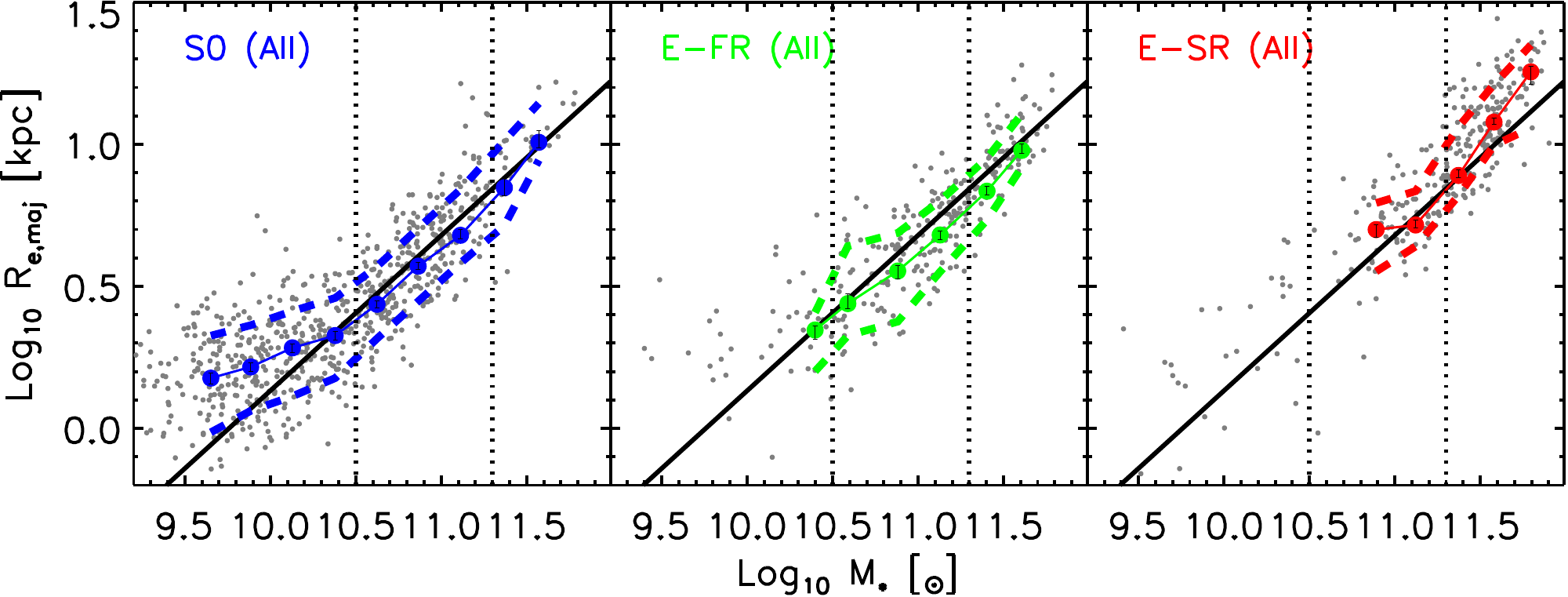}
  \includegraphics[width=0.8\linewidth]{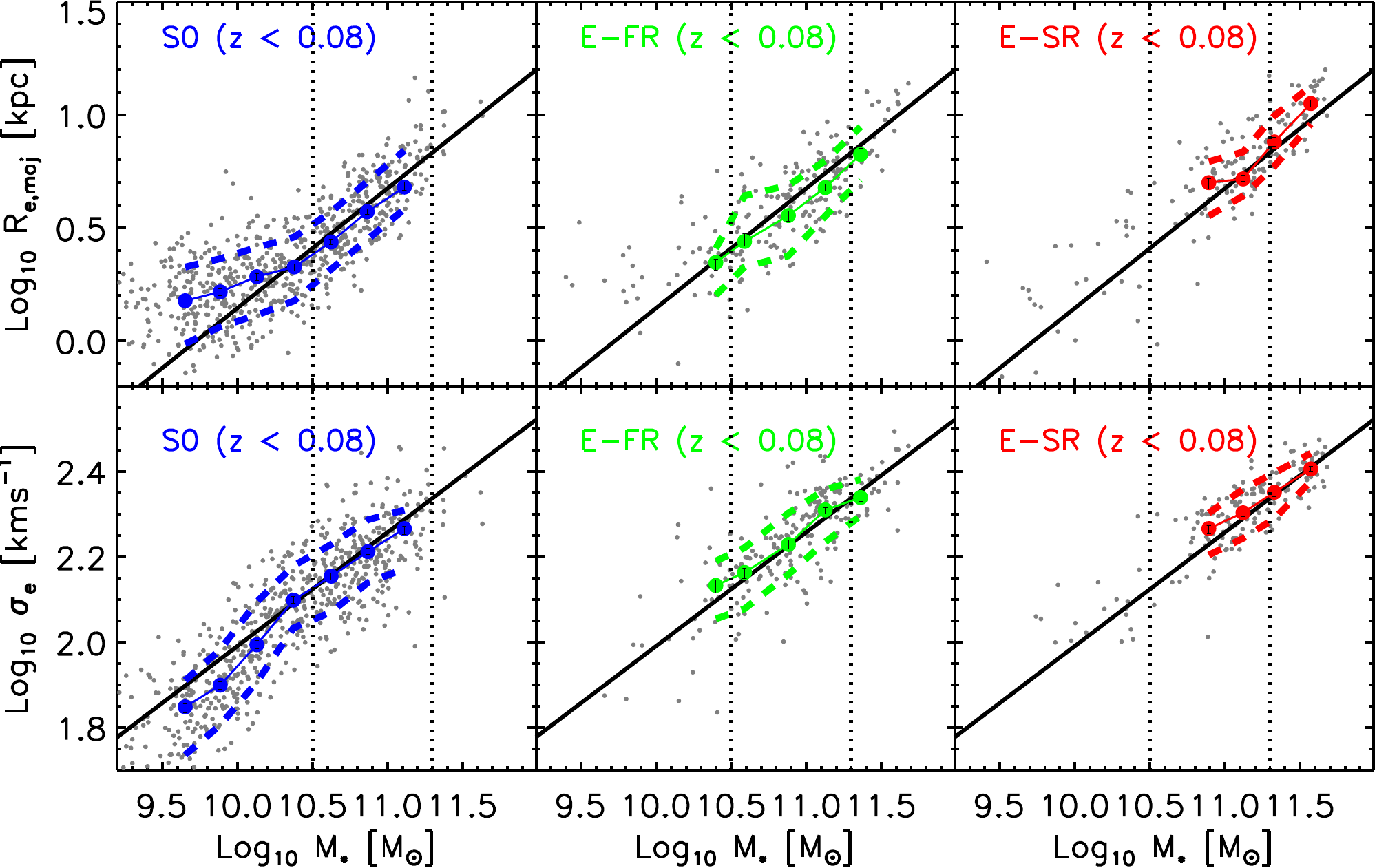}
  \caption{Correlation of stellar mass with size (top and middle) and velocity dispersion (bottom) for S0s (left), E-FRs with $\lambda_e>0.2$ (middle) and E-SRs (right) in MaNGA.  Middle panels show the subset of the objects in the top panels which have $z<0.08$.  Solid black lines in each panel show linear fits to the full sample (S0 + E-FR + E-SR) in the mass range $10.5\le \log_{10}M_*/M_\odot\le 11.3$, while colored curves show the median and quartiles of each subsample.  Dotted vertical lines show two mass scales at which the scaling relations curve away from this linear fit; this curvature is thought to reflect changes in the typical formation or assembly history.  Above $2\times10^{11}M_\odot$ dry mergers are expected to dominate; here, the fraction of S0s drops sharply and Es dominate the counts (Table~\ref{tab:frac}).  Below $3\times10^{10}M_\odot$ S0s dominate.  The remainder of this paper studies if the stellar populations and/or assembly histories of S0s change dramatically around $3\times10^{10}M_\odot$.}
  \label{fig:VL}
\end{figure*}

\subsection{Sample selection: Morphology}\label{sec:morph}
This paper uses a clean sample of S0 galaxies to study their properties as a function of global parameters (i.e. absolute magnitude, central velocity dispersion) as well as galactocentric distance (i.e. their gradients). 

To select our S0 sample we use the Deep Learning Morphological Value Added Catalog \citep[MDLM-VAC]{Fischer2019} that provides morphological properties derived from supervised Deep Learning algorithms based on Convolutional Neural Networks (see \citealt{DS2018} for more details).
The selection requires 
$$
 {\rm TType}\le 0   \qquad \textrm{and} \qquad  P_{\rm S0} >  0.5.
$$ 
The first condition selects early-type galaxies (1948 observations of 1908 galaxies) instead of late-type galaxies, and the second selects S0s (896 observations of 880 galaxies) rather than Es. About $\sim 20$\% of the  MaNGA DR15 galaxies are S0s.  Some previous work uses results from the Galaxy Zoo2 analysis \citep{Willett2013,Hart2016} to select lenticulars.  This can result in a sample in which Spirals contribute as much as 40\% \citep[e.g. see Figs.24-26 in][]{Fischer2019}.  In contrast, objects identified by the MDLM-VAC as S0s are significantly less likely to be Spirals (also see Section~\ref{sec:previous}). To illustrate, we show a random selection of images in the Appendix. 


In addition, the IFU observations provided by MaNGA enable an estimate of the angular momentum $\lambda_e$ defined in \citet{Emsellem2007} (corrected for seeing following \citealt{Graham2018}) for each galaxy.  When combined with the ellipticity $\epsilon$ returned by the MPP-VAC, this allows us to separate slow from fast rotating galaxies (e.g. Paper~II).  This $\lambda_e-\epsilon$ diagnostic shows that the vast majority of S0s are FRs (see Figs.~28--30 in \citealt[]{Fischer2019}).  While this is not surprising, it is reassuring, as the angular momentum played no role in the MDLM-VAC morphological classification.  Although, in principle, E-FRs can have $\lambda_e < 0.2$, in what follows, when we show E-FRs, they have $\lambda_e>0.2$ unless we specify otherwise.  This is similar to our convention in Papers~I and~II.

\begin{table}
 \centering
 MORPHOLOGICAL FRACTIONS AT FIXED $M_*$\\
 \begin{tabular}{ccccc}
 \hline
       $M_*$       &  S0 & E-FR & E-FR & E-SR \\
  $[10^{10}M_\odot]$ &     & $\lambda_e > 0.2$ & $\lambda_e < 0.2$ & \\
 \hline
  $ > 20$                  & $0.14$ & $0.28$ & $0.17$ & $0.41$ \\
  $3 - 20$ & $0.60$ & $0.23$ & $0.06$ & $0.11$\\
  $0.1 - 3$         & $0.87$ & $0.07$ &  $0.01$ & $0.05$\\
 \hline
 \hline
 \end{tabular}
 \caption{Morphological fractions in the specified mass bin, weighted by the MaNGA selection function, in each of three mass bins shown in Figure~\ref{fig:VL}.}
 \label{tab:frac}
\end{table}


\begin{table}
\centering
 $M_*$ DISTRIBUTION AT FIXED MORPHOLOGY\\
\begin{tabular}{ccccc}
 \hline
 Sample   & $M_* = 0.1 - 3$   &  $3 - 20$  & $ > 20$  \\
          & [$10^{10}M_\odot$] &  [$10^{10}M_\odot$] &  [$10^{10}M_\odot$] \\
 \hline
 Observed $z < 0.08$\\
 \hline
 E-SR &  0.10 & 0.48 &  0.42 \\
 E-FR ($\lambda_e < 0.2$) & 0.11  & 0.55 & 0.34\\
 E-FR ($\lambda_e > 0.2$) &  0.11 & 0.69 &  0.20 \\
 S0 &   0.48 & 0.49 &  0.03 \\
 \hline
 Observed All\\
 \hline
 E-SR &  0.06 & 0.28 & 0.66 \\
 E-FR ($\lambda_e < 0.2$) & 0.08  & 0.42 & 0.50\\
 E-FR ($\lambda_e > 0.2$) &  0.09 & 0.53 & 0.38 \\
 S0 &   0.44 & 0.49 & 0.07 \\
 \hline
 \hline
 Weighted \\
 \hline
 E-SR &  0.30 & 0.52 &  0.18 \\
 E-FR ($\lambda_e < 0.2$) & 0.27  & 0.60 & 0.13\\
 E-FR ($\lambda_e > 0.2$) &  0.24 & 0.68 &  0.08 \\
 S0 &   0.65 & 0.34 &  0.01 \\ 
 \hline
 \hline
\end{tabular}
\caption{Stellar mass distribution for the different morphological types in the $z<0.08$ sample, the full sample, and when objects have been weighted to account for the MaNGA selection function.}
\label{tab:fracMs}
\end{table}

\begin{figure}
  \centering
  \includegraphics[width=0.99\linewidth]{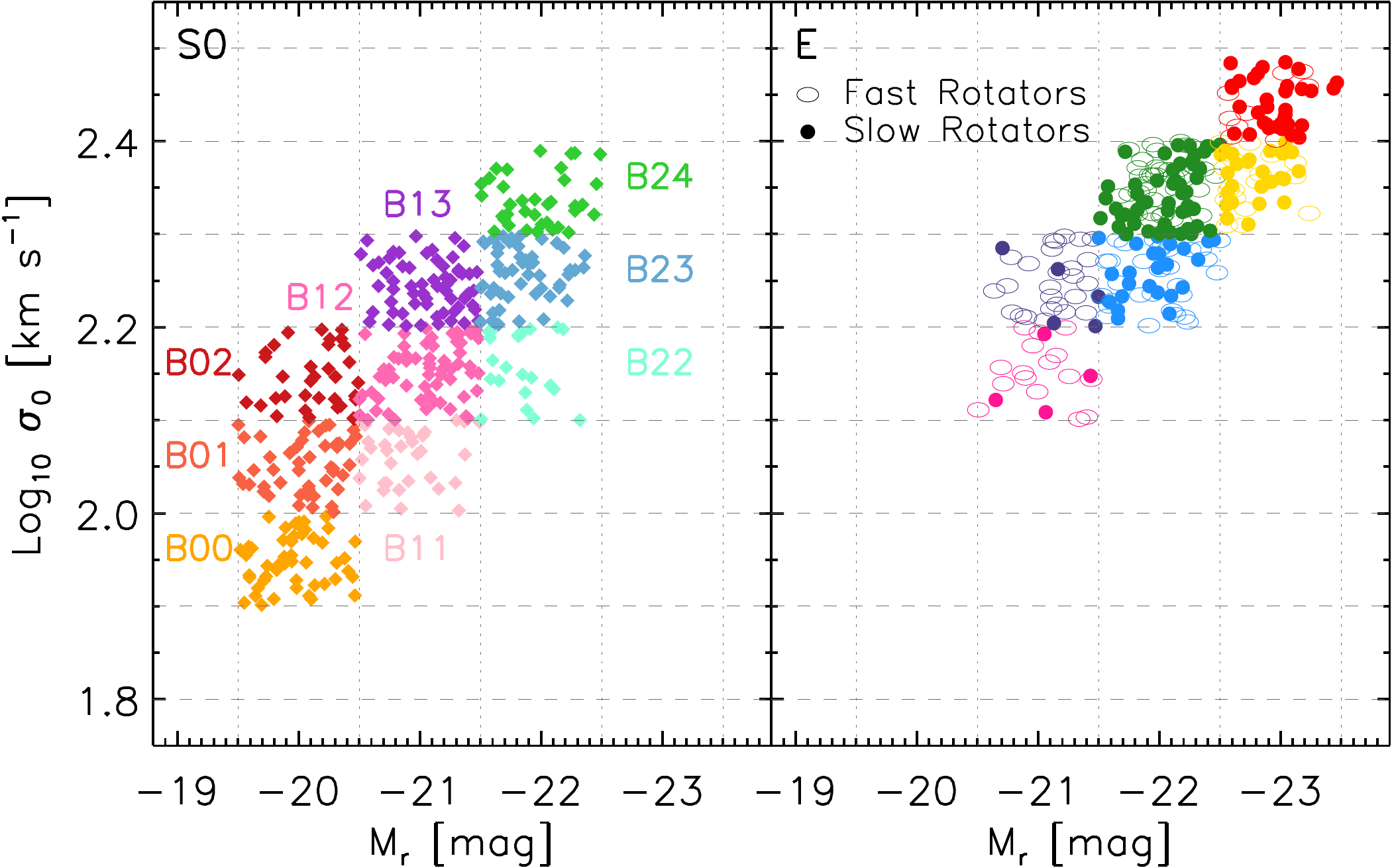}
  \caption{Distribution of central velocity dispersion $\sigma_0$ and absolute magnitude $M_r$ for S0s (left) and Es (right).  Filled and open ellipses in the right hand panel show slow and fast rotator Es.  FRs dominate at fainter luminosities; essentially all the S0s are FRs. Grid shows the values of $M_r$ and $\sigma_0$ which define the bins in (Table~\ref{tab:bin}).  The label B$ms$, where $m$ denotes the magnitude and $s$ the $\sigma_0$, indicates the bin reported in Table~\ref{tab:bin}.} \label{fig:sample}
\end{figure}

\begin{figure*}
  \centering
  \includegraphics[width=0.8\linewidth]{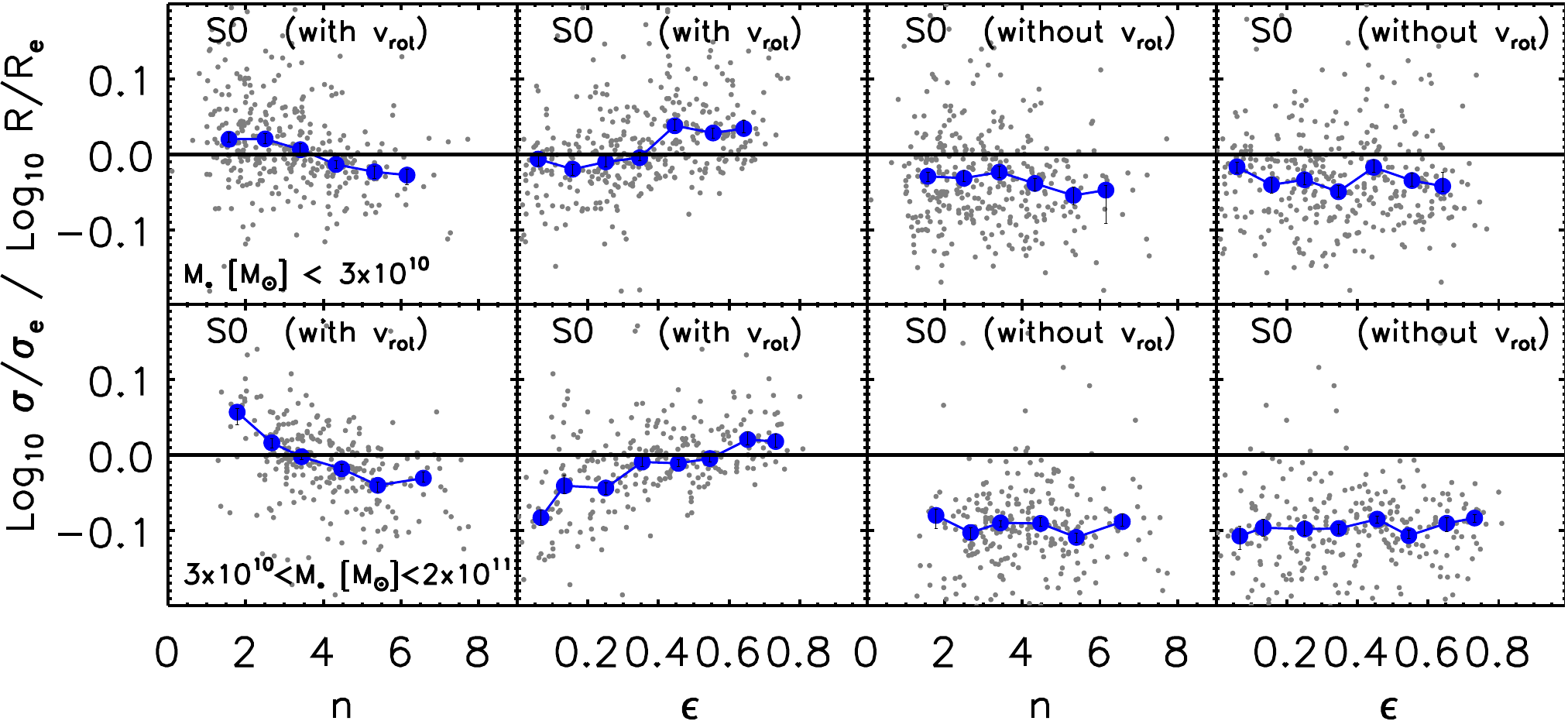}
  \caption{Velocity dispersion gradient for low (top) and high (bottom) mass S0s, shown as a function of Sersic index $n$ and ellipticity $\epsilon$.  In left hand panels, estimated dispersions include a contribution from rotation; in right hand panels, this rotation has been removed.  In all panels, symbols show individual galaxies and curves show the trend defined by the median value.} \label{fig:gradS}
\end{figure*}

\subsection{Curvature in scaling relations}\label{sec:curvature}
Before we analyse the stellar populations of our S0 sample, Figure~\ref{fig:VL} shows updated versions of plots which first appeared in \citet{Bernardi2011}, highlighting the presence of two mass scales in the early-type galaxy population:  $3\times 10^{10}M_\odot$ and $2\times10^{11}M_\odot$.  \cite[Following Paper~I, we define $M_*$ as the product of $M_*/L_r$ from][and $L_r$ from the MPP-VAC.]{Mendel2014}  The difference here is that we have used the MDLM-VAC classifications to separate the early-type population into S0s (left), E-FRs (with $\lambda_e > 0.2$; middle) and E-SRs (right). The top and middle panels show the $R_e$-$M_*$ correlation; they only differ because of a redshift cut. This redshift cut matters more for the E-SRs than for the S0s, which are the subject of this paper.  We have included the top panel since there are too few E-SRs at $z<0.08$ to see the change in curvature at $2\times10^{11}M_\odot$ clearly. The bottom panels show the $\sigma_e$-$M_*$ relation for the objects at $z<0.08$.  The vertical dotted lines identify the mass scales at which the scaling relations change.  

The MaNGA selection algorithm is complicated, so Figure~\ref{fig:VL} and the following figures show results when galaxies have been weighted by {\tt ESWEIGHT} to account for the MaNGA selection. We note that since most of our analysis is done in narrow mass bins the effect of weighting is negligible except for when we compare the observed counts in one mass bin with those in another (see Table~\ref{tab:fracMs}).

Table~\ref{tab:frac} shows the morphological fractions in fixed mass bins, weighted by {\tt ESWEIGHT}. These weighted counts show that S0s account for more than 85\% of the sample below $3\times10^{10}M_\odot$, but less than 15\% above $2\times10^{11}M_\odot$.  In contrast, E-SRs are only about 5\% of the objects below $3\times10^{10}M_\odot$ but about 40\% of the objects above $2\times10^{11}M_\odot$.  (These fractions do not depend on the redshift cut.)

While it is tempting to attribute some of the change in scaling relations at these mass scales to morphology, it is important to note that, even in the S0 population, the scaling relations change dramatically at $3\times10^{10}M_\odot$.  Papers~I and~II studied the E-FR and E-SR galaxies.  In what follows, we will perform a similar study of the S0s, paying particular attention to SPs and gradients above and below $3\times10^{10}M_\odot$.

\begin{table}
\centering
BINNING OF GALAXIES\\
\begin{tabular}{ccccc}
  \hline
  Bin &  M$_r$ & Log$_{10}$ $\sigma_0$ & S0 Galaxies & S0 Galaxies \\
      &  [mag] &     [km s$^{-1}$]    & $z\le 0.08$   &    all $z$     \\  
 \hline
 B00    & $-19.5, -20.5$  &  1.90, 2.00  & 50  & 50 \\
 B01    & $-19.5, -20.5$  &  2.00, 2.10  & 49 &  49 \\
 B02    & $-19.5, -20.5$  &  2.10, 2.20  & 36 &  36 \\
  \hline
 B11    & $-20.5, -21.5$  &  2.00, 2.10  & 35  &  35 \\
 B12    & $-20.5, -21.5$  &  2.10, 2.20  & 63 &  64 \\
 B13    & $-20.5, -21.5$  &  2.20, 2.30  & 63 &  63 \\
  \hline
 B22    & $-21.5, -22.5$  &  2.10, 2.20  & 22  &  23 \\
 B23    & $-21.5, -22.5$  &  2.20, 2.30  & 55 &  63 \\
 B24    & $-21.5, -22.5$  &  2.30, 2.40  & 37 &  39 \\
 \hline
 \hline
 \end{tabular}
 \caption{Bin definitions and number of S0s in each bin, with and without redshift cut.}
 \label{tab:bin}
\end{table}

Before moving on, Table~\ref{tab:fracMs} shows the range of stellar masses $M_*$ spanned by the different morphological types.  Here, the weighted fractions are the most relevant:  they show that only about a third of E-SRs have $M_*<3\times 10^{10}M_\odot$, whereas nearly two-thirds of S0s have such masses.  In addition, about 20\% of E-SRs but essentially no S0s have $M_*>2\times 10^{11}M_\odot$.  

\subsection{Sample selection: Binning in $\sigma_0$ and $L_r$}\label{sec:bins}

We remove from our sample galaxies with no available photometric parameters (FLAG$\_$FIT=3 in MPP-VAC), or with unreliable spectra due to contamination by neighbours (identified by visual inspection). We also limited our sample to galaxies with $z\le 0.08$ and R$_e$ $<$ 15 arcsec, for reasons we discuss in Paper~I. Note that the redshift cut has little impact: it only removes 12 S0 galaxies, 11 of which have M$_r$ $<$ -21.5 (see Table \ref{tab:bin}).

To measure absorption features accurately, in particular the TiO indices associated with IMF variations, spectra must have SN $\ge 100$.  As we discussed in Paper~I, this requires to create stacked spectra.  We do so by first binning in central velocity dispersion $\sigma_0$ and absolute magnitude $M_r$.  The bin limits and the number of galaxies in each bin are given in Table \ref{tab:bin}. The left hand panel of Figure~\ref{fig:sample} shows how the S0s (colored symbols) populate the 9 bins (grids).  For comparison, the right hand panel shows the distribution of slow and fast rotating Es.  Papers~I and~II considered the objects in the four brightest bins in the right hand panel.  Although our main focus here is on the S0s, in Section~\ref{sec:E+S0s} we compare S0s with E-FRs.

\section{Gradients in the S0 sample}\label{sec:grads}
\subsection{Velocity dispersion}\label{sec:sigma}

To measure SP gradients we follow the same methodology as in Paper~I (see section 2.3 there).  Briefly, for each $M_r$ and $\sigma_0$ bin, we compute stacked spectra from concentric ellipsoidal shells that have width $0.1R_{e,{\rm maj}}$ along the major axis (so they are $b/a$ thinner along the minor axis).  The stacked value is the median rest-frame normalized flux of all the individual spaxels from MAPS-VOR10-GAU-MILESHC with S/N $>$ 5 which belong to each radial bin. To avoid variations due to the spectral slope, the normalization is done in the continuum region of each of the lick indices that we measure (defined in Table 3 of Paper~I). The spectra are then smoothed to a common resolution of 300 km s$^{-1}$ to allow for a fair comparison between the different bins and the values from stellar population synthesis models.

To bring them to the 300 km s$^{-1}$ resolution, we first need to measure the velocity dispersion of each stack.  The velocity dispersion varies with galactocentric distance.  We illustrate this gradient by taking the ratio of $\log_{10}[\sigma(<R_{\rm min})/\sigma(<R_{\rm max})]$ to $\log_{10}(R_{\rm min}/R_{\rm max})$ (Figure~\ref{fig:gradS}). Where possible, we set $R_{\rm min} = R_e/4$ and $R_{\rm max}=R_e$.  However, for some galaxies, the angular size corresponding to $R_e/4$ is less than 1.2 arcsec, so seeing is an issue.   For these we set $R_{\rm min}=R_e/2$.  For a few other galaxies, the available spectra do not extend all the way out to $R_e$.  For these, we set $R_{\rm max}=3R_e/4$. Our notation $\sigma(<R)$ means we estimate $\sigma$ from a stacked spectrum constructed using all the available spaxels within $R$.  We can do this in two ways:  either by simply stacking all the spaxels directly (as this approximates what would have been observed by a large fiber), or by first removing the mean rotation speed from each spaxel before stacking.

Figure~\ref{fig:gradS} shows the gradients measured in this way as a function of Sersic index $n$ and image ellipticity $\epsilon$ (a face-on disk would have $\epsilon=0$) for low (top) and high (bottom) mass S0s.  There are clear trends with both $n$ and $\epsilon$ when rotation is included in the dispersion estimate (left hand panels).  These trends are not present when rotation has been removed (right hand panels).  Clearly, removing rotation is preferable.  Comparison of the top-right and bottom-right panels shows that, once rotation has been removed, the more massive S0s clearly have larger gradients.  Thus, from the velocity dispersion profiles alone, we already have a hint that the mass scale $3\times10^{10}M_\odot$, so apparent in Figure~\ref{fig:VL}, is special.  

\begin{figure}
  \centering
  \includegraphics[width=0.9\linewidth]{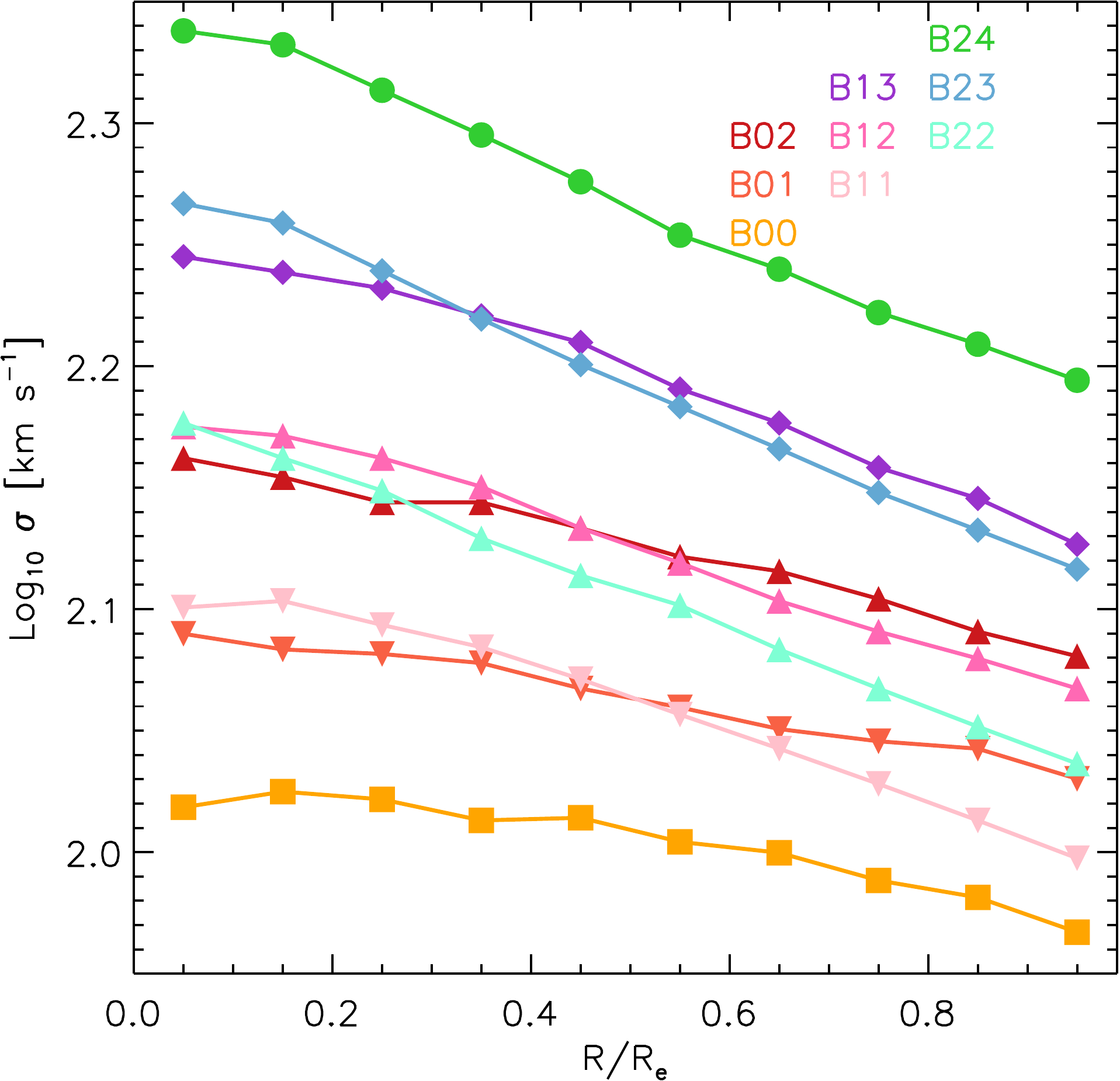}
  \caption{Velocity dispersion profiles obtained from the stacked spectra of the S0 sample. Colors correspond to the bins presented in Figure \ref{fig:sample}, as stated in the legend (the same symbols are used for bins with the same $\sigma_0$).  S0s with larger $\sigma_0$ have steeper profiles and, for the same $\sigma_0$, more luminous galaxies have steeper profiles. } \label{fig:vdisp}
\end{figure}

\begin{figure*}
  \centering
  \includegraphics[width=0.6\linewidth]{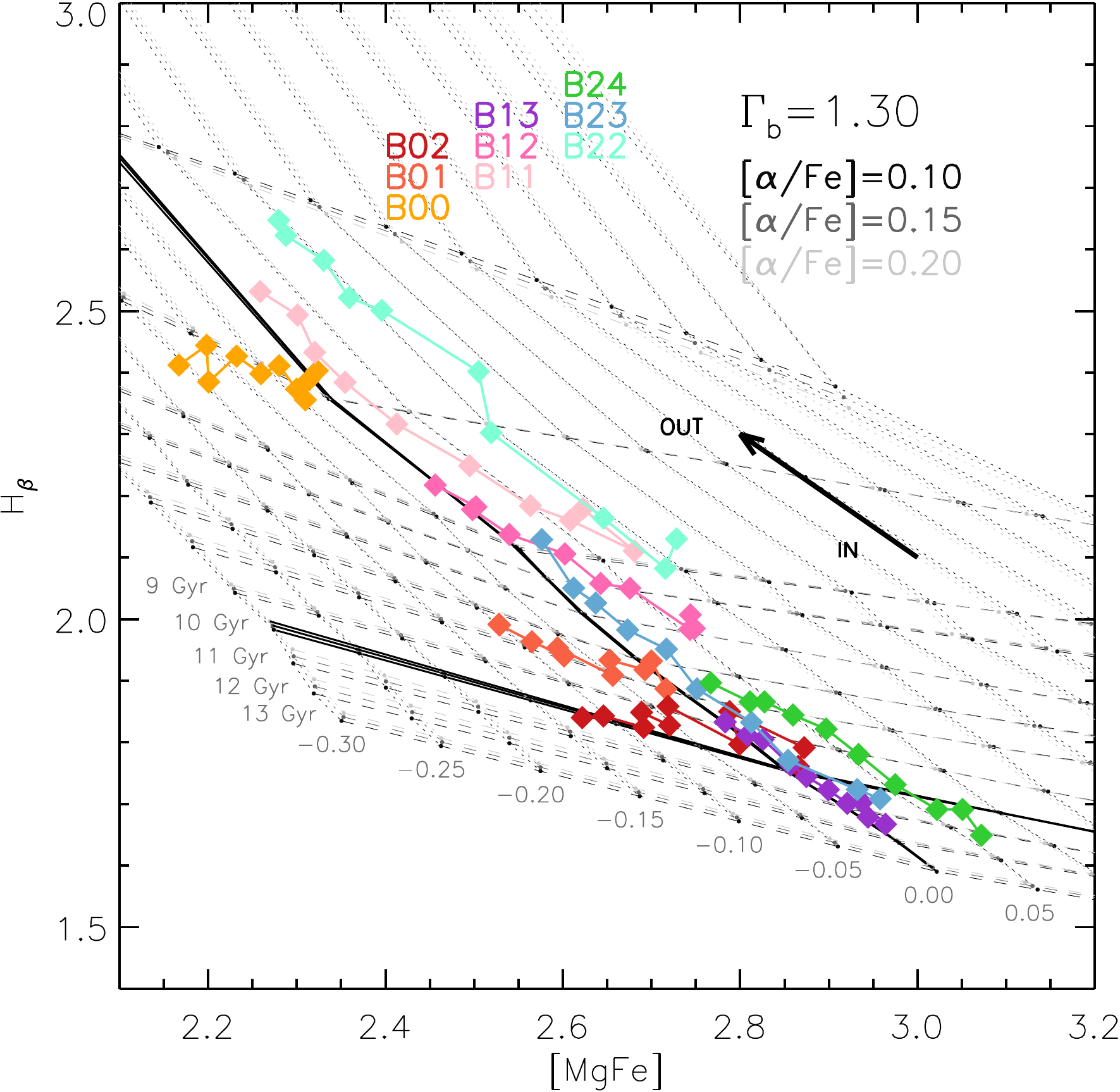}
  \caption{Lick index H$_{\beta}$ versus [MgFe] diagram for the S0 sample.  Colored symbols show measurements from the stacked spectra in each bin at different galactocentric distances as indicated by the arrow.  These are superimposed on age-metallicity grids from the MILES-Padova models with other SP parameters as indicated. Model grids all have same IMF slope ($\Gamma_b=1.3$) but different $\alpha$-enhancement values (shown as different grey scales -- the grids are nearly identical).} \label{fig:HbMg}
\end{figure*}
\begin{figure}
  \centering
  \includegraphics[width=0.9\linewidth]{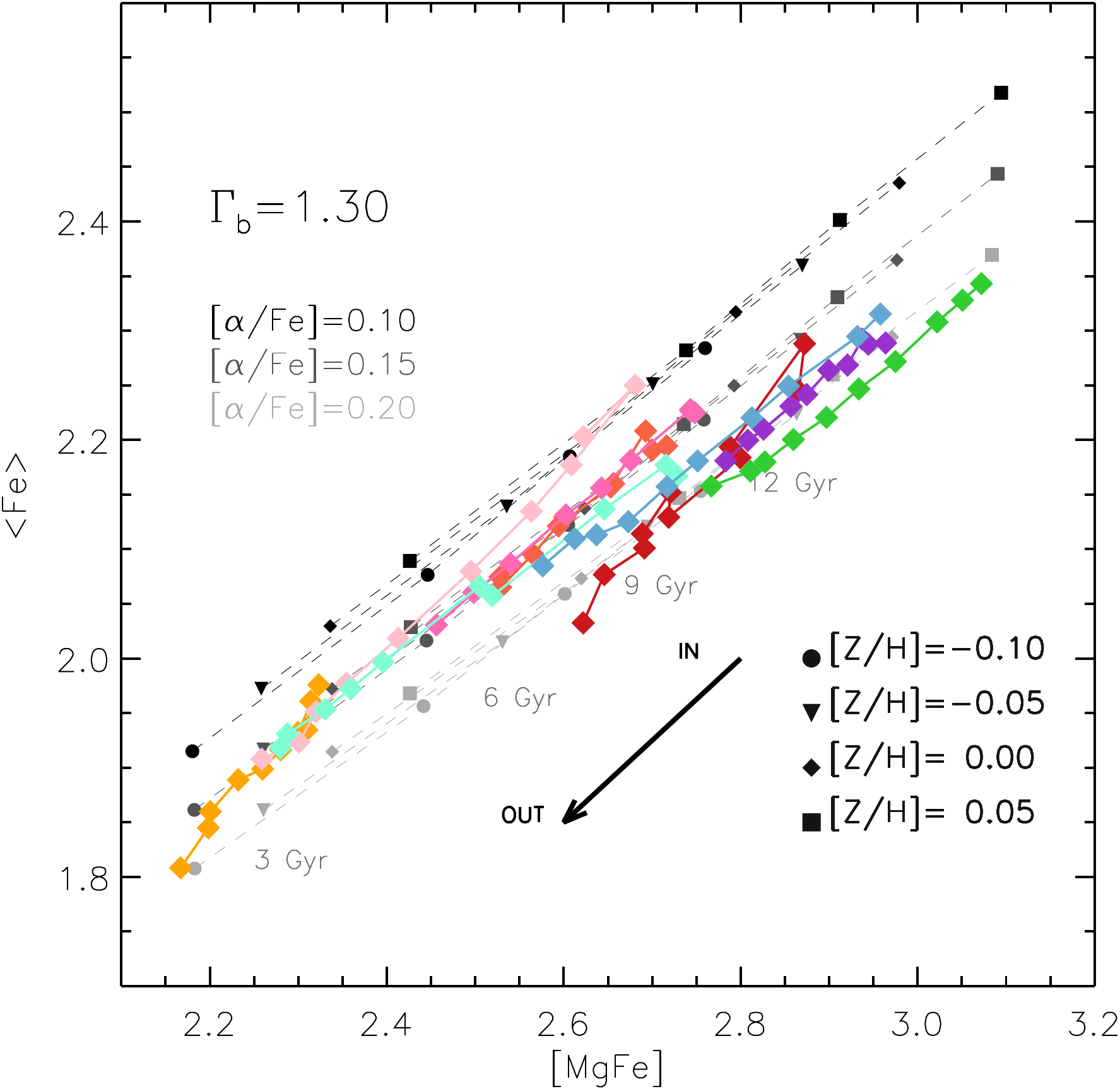}
  \caption{Same as Figure~\ref{fig:HbMg} but now for <Fe>.
    Model grids again all have $\Gamma_b=1.3$, but the dependence on $\alpha$-enhancement is much stronger.} \label{fig:Fe}
\end{figure}
\begin{figure}
  \centering
  \includegraphics[width=0.9\linewidth]{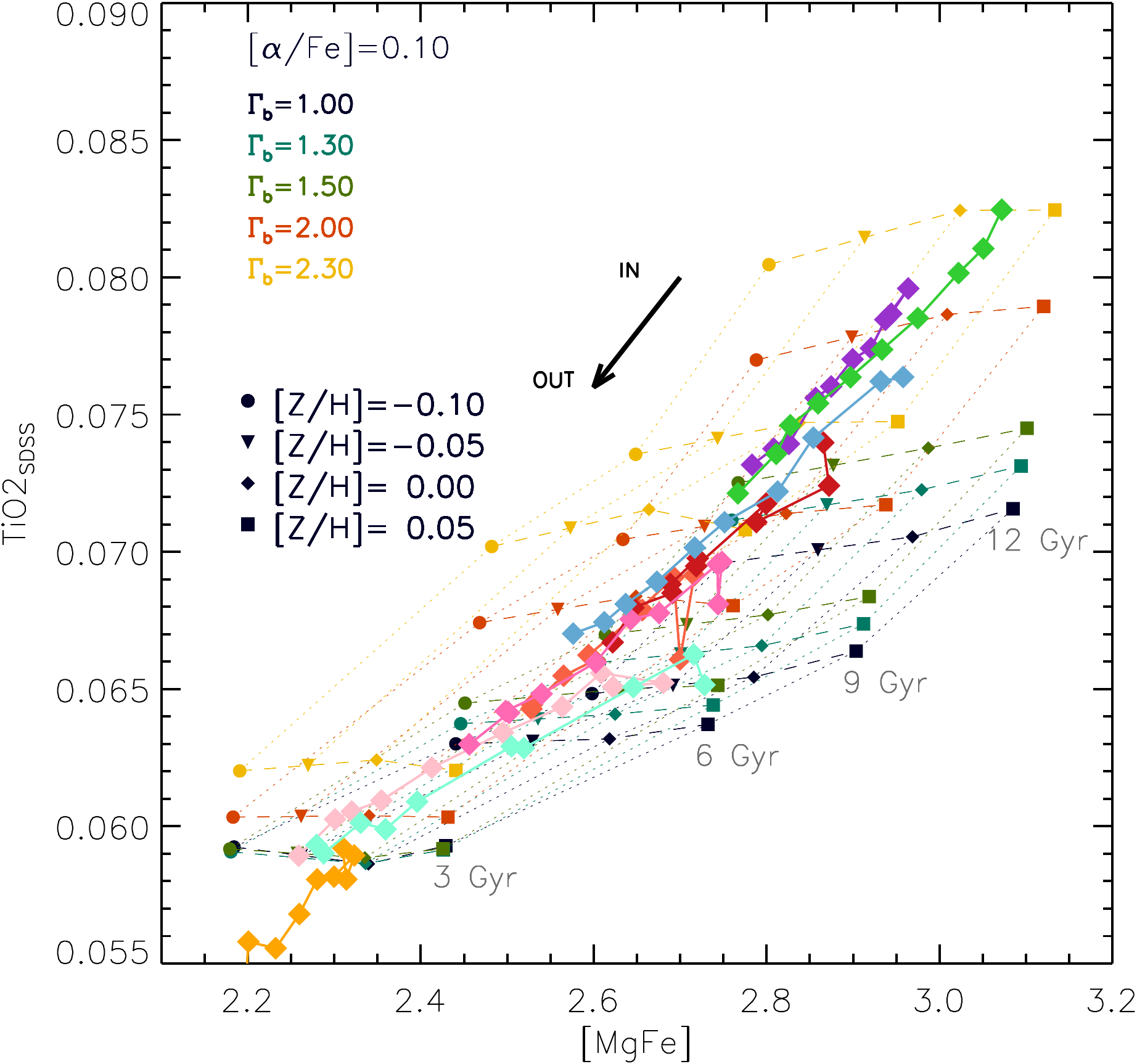}
  \caption{Same as Figure~\ref{fig:HbMg} but now for TiO2$_{\rm SDSS}$.
    In this case, [$\alpha$/Fe] is fixed to $0.10$ and grids with different IMF slopes are shown, color-coded according to the legend.} \label{fig:TiO2}
\end{figure}

Figure~\ref{fig:vdisp} shows the full velocity dispersion profile for each bin after removing the mean rotation speed from each spaxel before stacking. It is easy to see that the profiles are steep at larger $\sigma_0$ values and, at fixed $\sigma_0$, steeper for the brightest galaxies (compare the cyan with the dark red lines).  In fact, the fainter (M$_r> -20.5$) S0s show a much flatter velocity dispersion profile -- consistent with Figure~\ref{fig:gradS} (right panels) -- which, we will show in the next sections, comes along with a strong gradient in metallicity but almost no age gradient.

\subsection{Lick indices and SSP models}\label{sec:lick}
For each stack, we measure a set of Lick indices which we compare to the predictions of the MILES SSP models \citep{Vazdekis2010} to estimate age, metallicity, $\alpha$-enhancement, IMF slope and M$_*$/L. In particular, we use three index-index diagrams to constrain our parameter space: H$_{\beta}$-[MgFe] is sensitive to age and metallicity, <Fe>-[MgFe] is sensitive to the $\alpha$-enhancement and TiO2$_{\rm SDSS}$-[MgFe] is a good IMF indicator. The indices are defined following \cite{Trager1998}, except for TiO2$_{\rm SDSS}$ which was introduced by \cite{LaBarbera2013} (see Table~3 from Paper~I).

We use the MILES-Padova models with BiModal IMFs to interpret our measurements. Our results depend on a number of assumptions such as the choice of SSP models, the IMF parametrisation or the indices used as IMF indicators, as discussed in the Appendix of Paper~I. To ease the comparison with the results from Paper~I and II, we use exactly the same methodology (including SSP library, $\alpha$-enhancement correction of the MILES-Padova models and best-fitting procedure) explained in detail in Section~3 of Paper~I. In Paper~I we also discussed the impact of variations in [X/Fe] elements when constraining IMF slopes (Figure~12 from Paper~I) and we tested 4 different assumptions (Figure~13 and Table 4 from Paper~I). In this work we restrict our analysis to what we called assumption 3 (i.e., allowing for IMF variations for each bin and with galactocentric distance) and fixing  $\Delta_{[\rm X/Fe]}= 0.003$.

\begin{figure}
  \centering
  \includegraphics[width=0.9\linewidth]{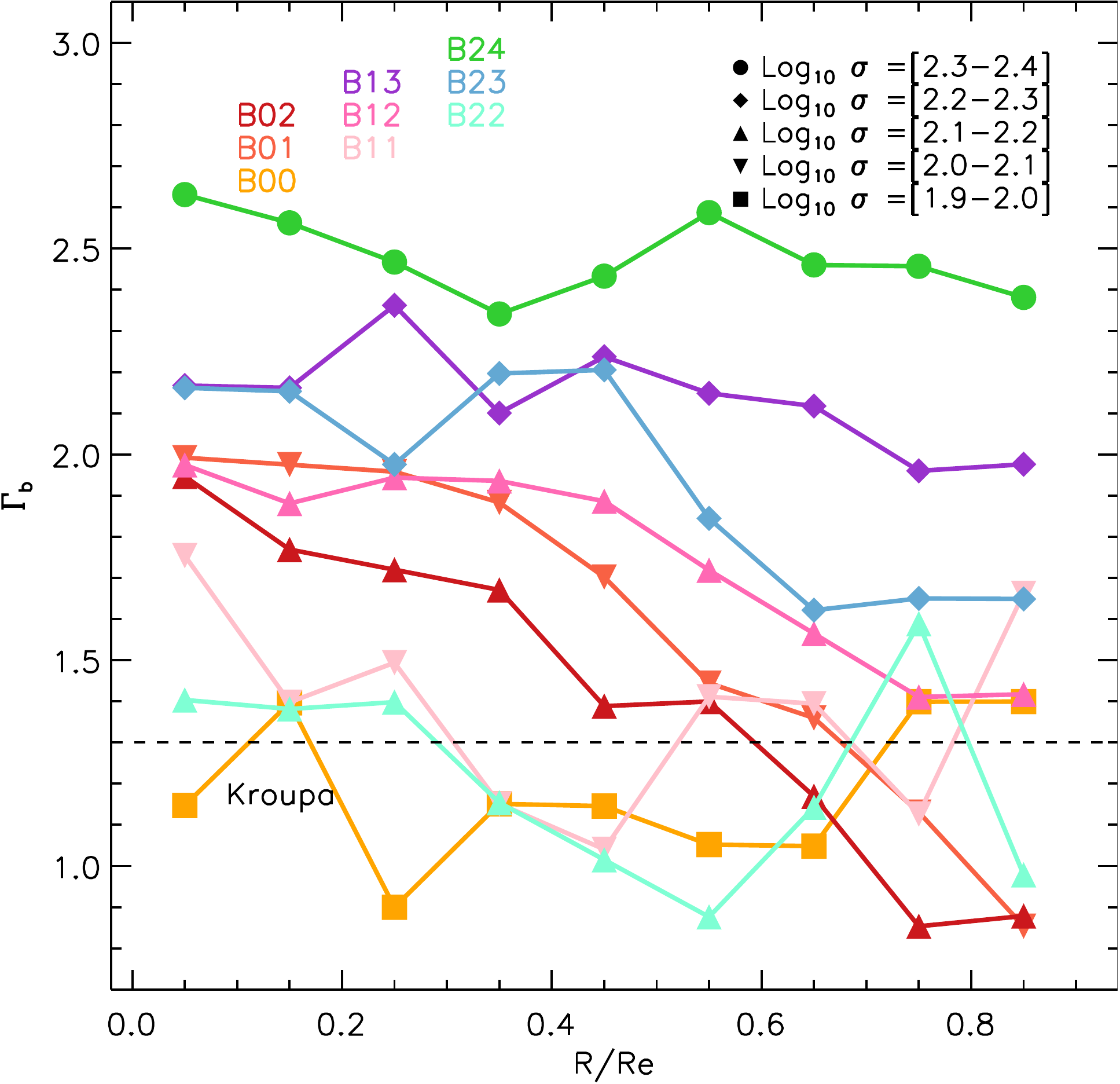}
  \caption{Best-fitting IMF slope as a function of galactocentric distance. Colors represent the bins defined in Figure~\ref{fig:sample}; symbols indicate the value of $\sigma_0$. The IMF slope is estimated by comparing the Lick indices, especially TiO2$_{\rm SDSS}$, with the MILES bimodal IMF SSP models. The horizontal dashed line indicates the Kroupa IMF.} \label{fig:IMF}
\end{figure}

Figure~\ref{fig:HbMg} shows how H$_{\beta}$ varies with [MgFe]:  Colored symbols show the values measured for each bin and grids show the MILES SP models. There are obvious differences between the radial gradients of the fainter galaxies (red, orange, yellow) and the brighter ones: galaxies with M$_r > -20.5$ span only a small range of H$_{\beta}$ whereas brighter galaxies span a much larger range of index strengths.  Thus, even without using SSP models to interpret the measurements, the scale M$_r \sim -20.5$ (which corresponds to a mass scale $M_* \sim 3\times10^{10}M_\odot$) is clearly special.

\begin{figure}
  \centering
   \includegraphics[width=0.9\linewidth]{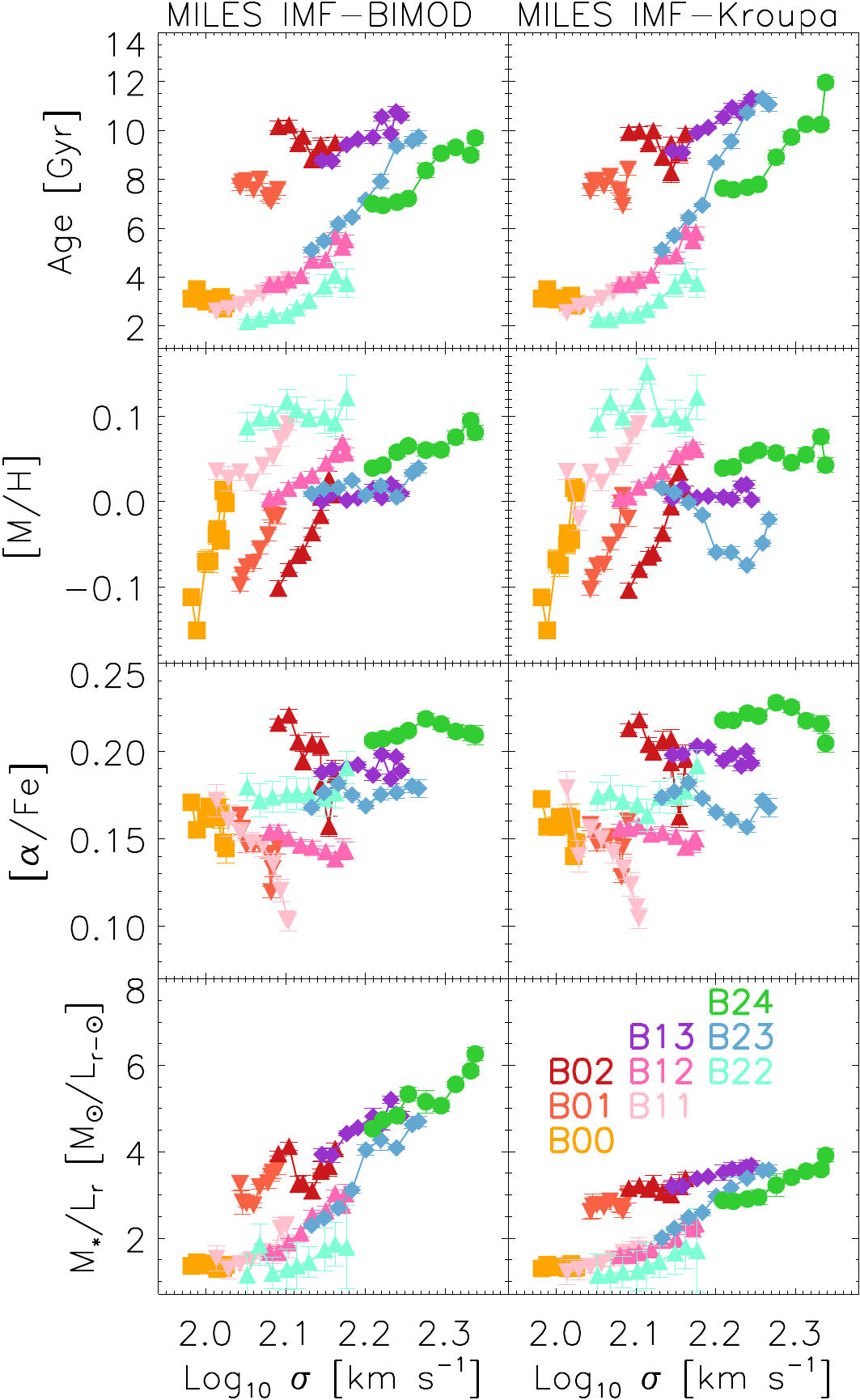}
  \caption{Best-fitting age, metallicity, $\alpha$-enhancement and M$_*$/L$_r$  versus $\sigma$ when allowing for IMF variations (left) and when fixing the IMF to Kroupa (right). Colors represent the different bins as stated in the legend, and symbols represent the $\sigma_0$ of each bin - as in Figure \ref{fig:IMF}. Error bars show uncertainties obtained by bootstrapping galaxies in each bin (see text). } \label{fig:results-sig}
\end{figure}

Figures~\ref{fig:Fe} and~\ref{fig:TiO2} show a similar comparison of the <Fe>-[MgFe] and TiO2$_{\rm SDSS}$-[MgFe] relations defined by the different bins.  There are again obvious trends with $\sigma_0$ at fixed $L_r$:  e.g., the <Fe>-[MgFe] relations shift to larger [MgFe] as $\sigma_0$ increases and TiO2$_{\rm SDSS}$ increases with $\sigma_0$ in each $L_r$ bin.  

\begin{figure*}
  \centering
   \includegraphics[width=0.9\linewidth]{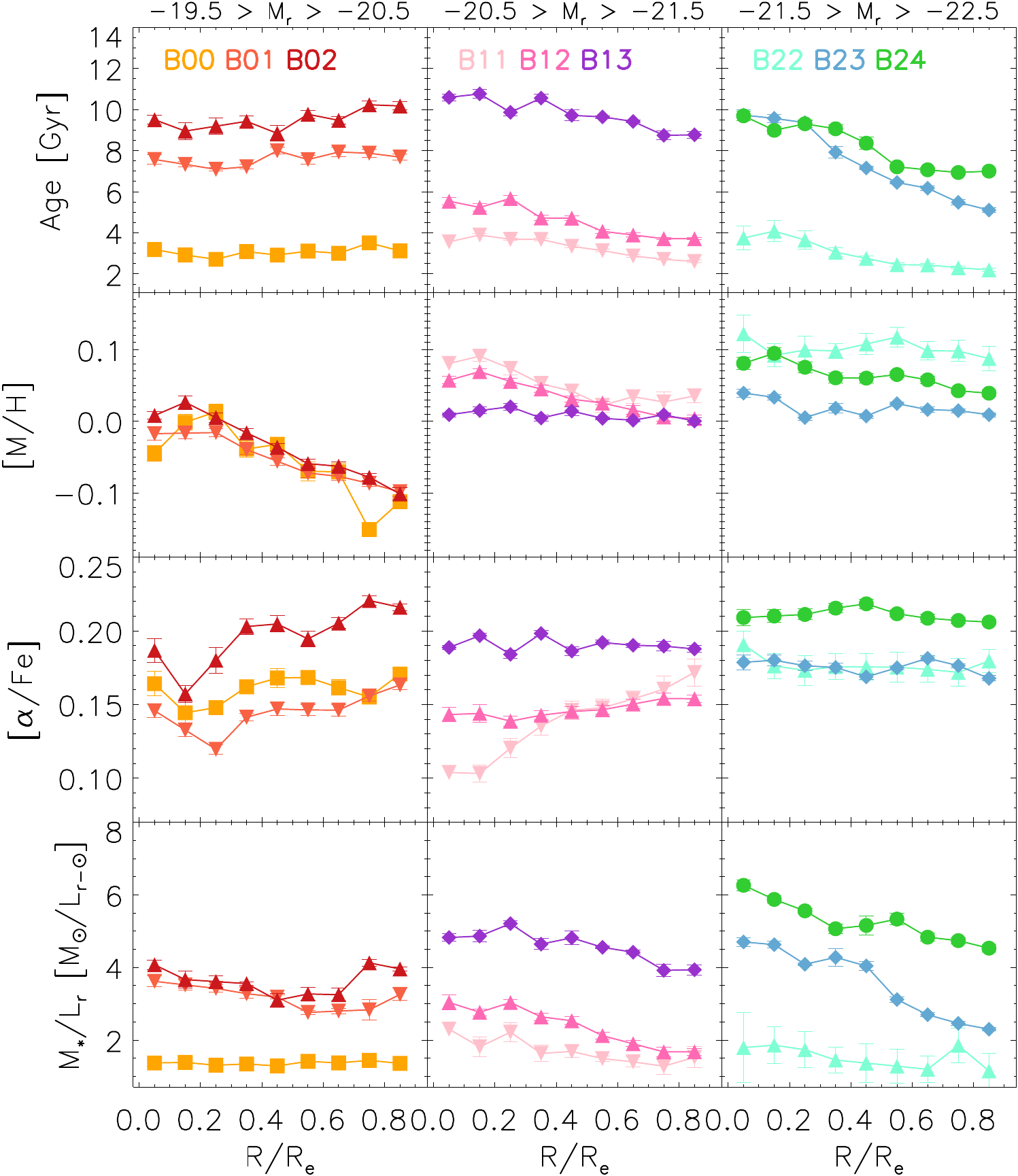} 
  \caption{Age, metallicity, $\alpha$-enhancement and M$_*$/L versus R/R$_e$, when the IMF is allowed to vary. Each column shows bins in the same luminosity range. Colors and symbols as in Figure \ref{fig:results-sig}.} \label{fig:results-rad}
\end{figure*}

Comparison with the model grids in Figure~\ref{fig:HbMg} shows that faint galaxies show almost no age gradient and, while having mainly sub-solar metallicities, show a large metallicity gradient.  In contrast, galaxies with M$_r < -20.5$ show a stronger age gradient and little if any metallicity gradient.  This is our first main result.  In addition, the grids show that, at fixed $M_r$, age increases with $\sigma_0$ (orange is younger than red, pink younger than purple, cyan younger than green).  Finally, the grids in Figure~\ref{fig:Fe} show that, at fixed $L_r$, [$\alpha/$Fe] increases with $\sigma_0$, and those in Figure~\ref{fig:TiO2} suggest that the IMF slope steepens with $\sigma_0$.  We use the MILES SSP models to quantify these trends in the next subsection.  

\subsection{Stellar population gradients}\label{sec:SPG}
As we did in Papers~I and~II, we allow for variations in the IMF slope when fitting the other SP parameters. The best-fitting IMF-slope for each bin is shown in Figure \ref{fig:IMF}. There are significant differences in the IMF slope for the different populations, ranging from bottom heavy IMF ($\Gamma_b$=2.5) for galaxies with the largest $L_r$ and $\sigma_0$ to Kroupa-like values \citep[$\Gamma_b$=1.3]{Kroupa2001} for the other extreme (smallest $L_r$ and $\sigma_0$). Although the radial gradients are noisy, there is a clear trend for IMF slope to decrease with decreasing $\sigma_0$ (at fixed $L_r$). However, galaxies with the same $\sigma_0$ do not share the same IMF slope, implying that luminosity also correlates with the IMF.  Finally, IMF gradients are obvious for some bins (e.g., B01, B02, B12) and almost negligible for others (e.g., B00, B22).  

Figure \ref{fig:results-sig} compares the best-fitting SP parameters as a function of velocity dispersion when the IMF is allowed to vary and when it is fixed (to Kroupa). The IMF variations have an important effect in the inferred M$_*$/L$_r$ values, which affect both stellar mass estimates -- M$_{\rm dyn}$ and M$_*$ -- as we stressed in Papers~I and~II.  However, the effect of the IMF on the other SP parameters is much weaker, and the overall shape and relative values between bins remain almost unchanged. This figure highlights the relation between the velocity dispersion gradient and those in age and metallicity: the low luminosity bins (red, orange and yellow) show relatively flat velocity and age profiles, while in the other bins the velocity and age gradients are stronger.  The metallicity gradients are strongest for the lowest luminosity bins, weaker for intermediate luminosities, and almost flat for the most luminous ones.  We discuss these trends in Section~\ref{sec:discuss}.

Figure \ref{fig:results-rad} shows how the SP estimates shown in the left hand panels of Figure~\ref{fig:results-sig} (i.e. when allowing for IMF variations) change with galactocentric distance (rather than $\sigma$). Except for M$_*$/L$_r$, fixing the IMF to Kroupa makes little difference.)  To simplify the comparison, bins of different luminosities are plotted separately in each column. It is evident that, at fixed $M_r$, age increases with $\sigma_0$ while metallicity decreases (except at $M_r > -20.5$, where the values are very similar). Stated differently, at fixed $\sigma_0$, more luminous galaxies are younger and more metal rich (compare upright triangles in the three top panels, or the upside-down triangles in the left and middle panels).  Evidently, the usual statement that `massive galaxies are older' (c.f. Introduction) is not true if $\sigma_0$ is held fixed.  

Regarding the $\alpha$-enhancement, at fixed luminosity, [$\alpha$/Fe] decreases with $\sigma_0$ (compare purple and light pink tracks), but become very similar for the lowest $\sigma_0$ bins at each M$_r$ (e.g., compare cyan and blue). The $\alpha$-enhancement profiles are almost flat for all the bins, except for bins B02 (red) and B11 (light pink), which show positive gradients. Finally, the bottom panels show that, at fixed $L$, M$_*$/L increases with $\sigma_0$ but gradients are stronger for the more luminous S0s.  

In Figures~\ref{fig:results-sig} and~\ref{fig:results-rad} the error bars show the 1$\sigma$ variation in the best-fitting parameters after bootstrapping one galaxy at a time when constructing the stacked spectra. These uncertainties are significantly smaller than any systematics due to the use of different SSP models,  IMF parametrizations, or IMF indicators. They are even smaller than the differences due to normalising the stacks in different spectral regions. Therefore, while we are very confident of our stacking procedure, we warn the reader about the other assumptions which can affect the results presented here.

\section{Are fast rotator ellipticals just face-on S0 galaxies?}\label{sec:E+S0s}
So far we have concentrated on the properties of S0s.  As we noted in the Introduction, whether or not S0s, which are almost exclusively fast rotators, are fundamentally different from fast rotating Es is an open question.  Recent work asserts that E-FRs are just S0s viewed face-on \citep[e.g.][]{C16, Graham2019}.  In this section we compare kinematic profiles and SPs of S0s and E-FRs of the same luminosity and velocity dispersion.  Comparison of the left and right hand panels of Figure~\ref{fig:sample} shows the region in the $M_r$-$\sigma_0$ plane where there is sufficient overlap to perform this comparison.  This is only possible for bins B12, B13, B23 and B24.

\begin{figure}
  \centering
  \includegraphics[width=0.99\linewidth]{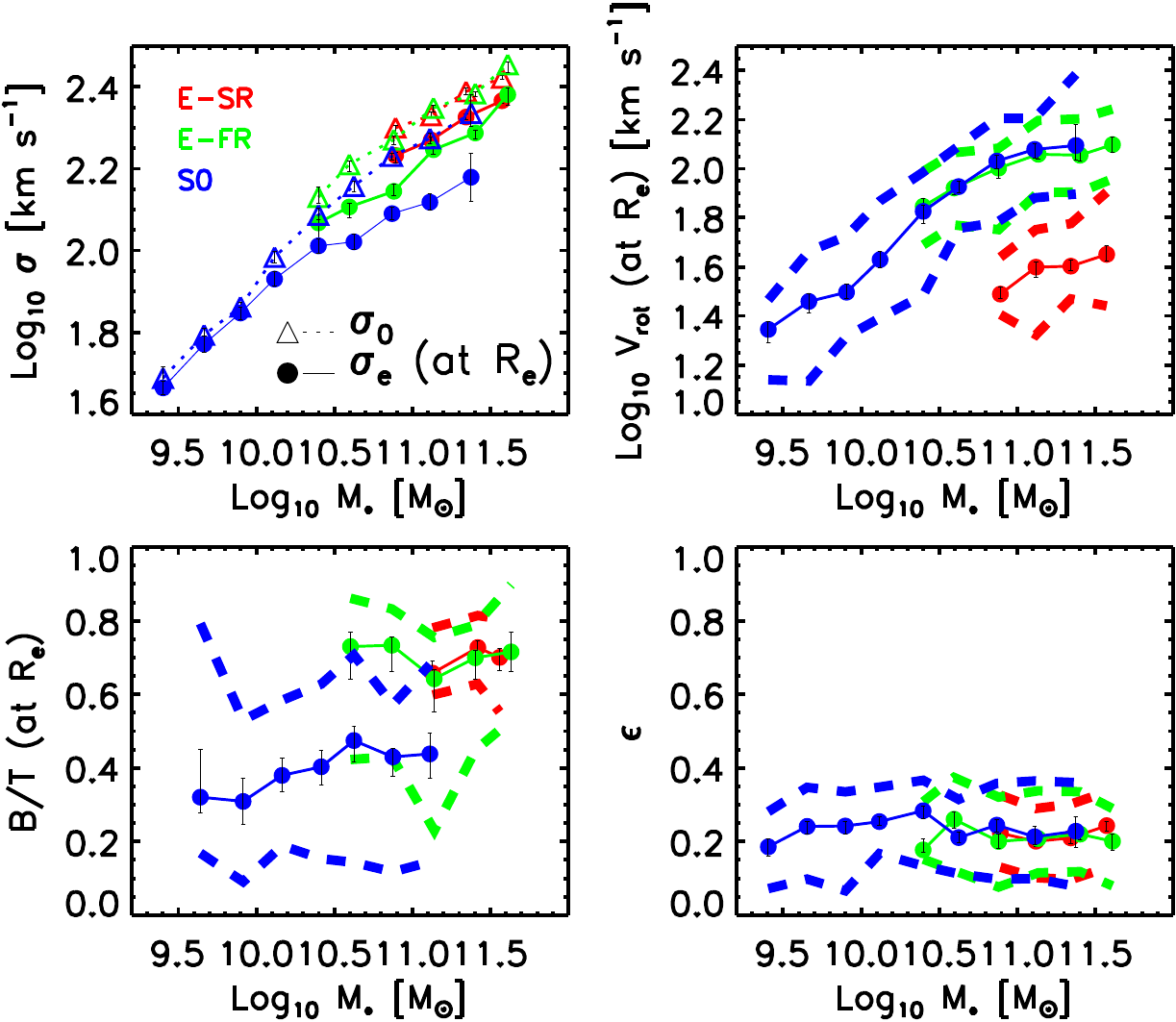}
  \caption{Velocity dispersion (central and at $R_e$), rotation speed, B/T and $\epsilon$ shown as median and quartiles as a function of stellar mass for E-SRs (red), E-FRs (green) and S0s with $\epsilon < 0.4$ (blue).} \label{fig:sigVrotBTe}
\end{figure}

\begin{figure}
  \centering
  \includegraphics[width=0.99\linewidth]{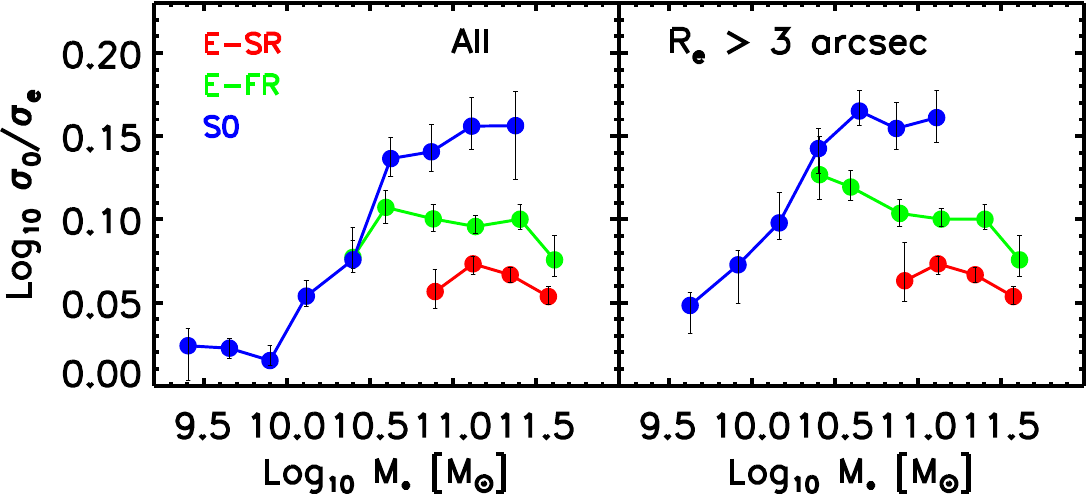}
  \caption{Ratio of central $\sigma_0$ to $\sigma_e$ (the values shown in the top left panel of the previous Figure) for S0s (blue) E-FRs (green) and E-SRs (red) as a function of stellar mass, for the full sample (left) and when restricted to objects with $R_e>3$~arcsec (right). } \label{fig:sig0sige}
\end{figure}

Whereas the reason for comparing the SPs is self-evident, our motivation for comparing kinematics deserves further comment.  Traditionally, S0s are thought to differ from Es because they have obvious inner dispersion supported bulge and outer rotation supported disc components.   In IFU datasets, this means that rotation should be obvious in the outer regions of S0s, provided they are not viewed face-on.  If face-on, then this rotation should be less obvious.  Moreover, the velocity dispersion in the outer parts, where the disc dominates, should be small, because we expect discs to be somewhat `cold'. In contrast, morphologically, E-FRs do not have an obvious disc component.  Therefore, there is no particular reason to expect the velocity dispersion profiles of E-FRs to differ from those of E-SRs.  Therefore, comparing the rotation and velocity dispersion profiles of E-FRs and face-on S0s of similar $L$ and $\sigma_0$ may be useful.

\subsection{Velocity dispersion profiles}
To explore this, we first select from our E and S0 samples only those objects which have $\epsilon<0.4$ (so face-on objects are a greater fraction).  Figure~\ref{fig:sigVrotBTe} shows how some kinematic (top panels) and photometric (bottom) properties of the objects which satisfy this cut vary with stellar mass.  The solid curves in the bottom right panel show that the median $\epsilon$ is similar for all three morphological types.  The associated dashed curves show that the quartiles around the median are also similar.  The bottom left panel shows that the E-SR and E-FR samples have similar median B/T values of about 0.7, whereas the S0s have a smaller median B/T, with a small but statistically significant step at $3\times 10^{10}M_\odot$. S0s with B/T values above their median overlap with E-FRs that lie below their median. The top right panel shows that E-FRs and S0s have very similar rotation speeds which are, of course, much larger than for E-SRs.  Finally, the top left panel shows that E-FRs have central velocity dispersion $\sigma_0$ similar to E-SRs and slightly larger than S0s of the same $M_*$.  The difference between S0s and the Es is slightly larger for $\sigma_e$ (at, not within, $R_e$), in qualitative agreement with the discussion in the previous paragraph.  (Rotation does not contribute to either of the estimated $\sigma$s.)

As we noted above, we are particularly interested in the slope of the velocity dispersion profile.  Figure~\ref{fig:sig0sige} shows a crude measure of this slope -- the ratio $\sigma_0/\sigma_e$ -- for the three morphological types, as a function of $M_*$.  At high masses, where a comparison is possible, this ratio is largest for S0s and smallest for E-SRs.  Evidently, the velocity dispersion drops more steeply for (face-on) S0s than it does for Es.

To check if the fiber size (radius $\sim 1$ arcsec) is compromising our results, the right-hand panel shows the result of restricting the analysis to objects having $R_e>3$~arcsecs, for which the aperture of the fiber should matter less.  Comparison of the two panels shows that, while there are some differences at smaller masses, over the range where all the morphological types can be compared, the differences are robust. (Restricting to $R_e>4$~arcsecs reduces the sample size considerably but does not change our conclusions.) The steeper $\sigma$ profiles for S0s, despite similar rotation speeds (see top right panel of Figure~\ref{fig:sigVrotBTe}), indicate that S0s tend to be more rotationally supported than are E-FRs.

It is worth noting that, for S0s, both B/T and the $\sigma_0/\sigma_e$ ratio change dramatically around $3 \times 10^{10}M_\odot$, the same mass scale where their stellar population gradients change.  Some of the drop and leveling-off in $\sigma_0/\sigma_e$ at low $M_*$ may be artificial, because both values are close to the instrumental resolution of the spectrograph ($\sim 60$~km/s, see top left panel of Figure~\ref{fig:sigVrotBTe}).  While this resolution has been accounted-for in our measurements, the $\sigma$ estimates become increasingly uncertain below this value.  Of course, this does not affect our conclusions about the differences between S0s and E-FRs at higher masses.

Before moving on to consider the stellar populations, it is natural to ask if the difference in velocity dispersion profiles is simply due to differences in B/T ratio.  The bottom left panel of Figure~\ref{fig:sigVrotBTe} shows that, at fixed $M_*$, S0s have smaller B/T than E-FRs, so one would expect S0s to have steeper velocity dispersion profiles.  However, the Figures show that at fixed morphology, the B/T$-M_*$ and $\sigma_0/\sigma_e-M_*$ correlations are both flat (above $3\times 10^{10}M_\odot$).  As a result, it is not obvious that B/T drives the differences we see in Figure~\ref{fig:sig0sige}.  Indeed, if we only select objects with $0.45<$ B/T $<0.6$ and $10.5 < \log_{10}(M_*/M_\odot) < 11$ then S0s and E-FRs have $\log_{10}(\sigma_0/\sigma_e)\approx 0.1$ and 0.15, just as in Figure~\ref{fig:sig0sige}.  Unfortunately, there are so few objects in this bin that it is hard to make a more statistically significant statement.

\begin{figure}
  \centering
  \includegraphics[width=.98\linewidth]{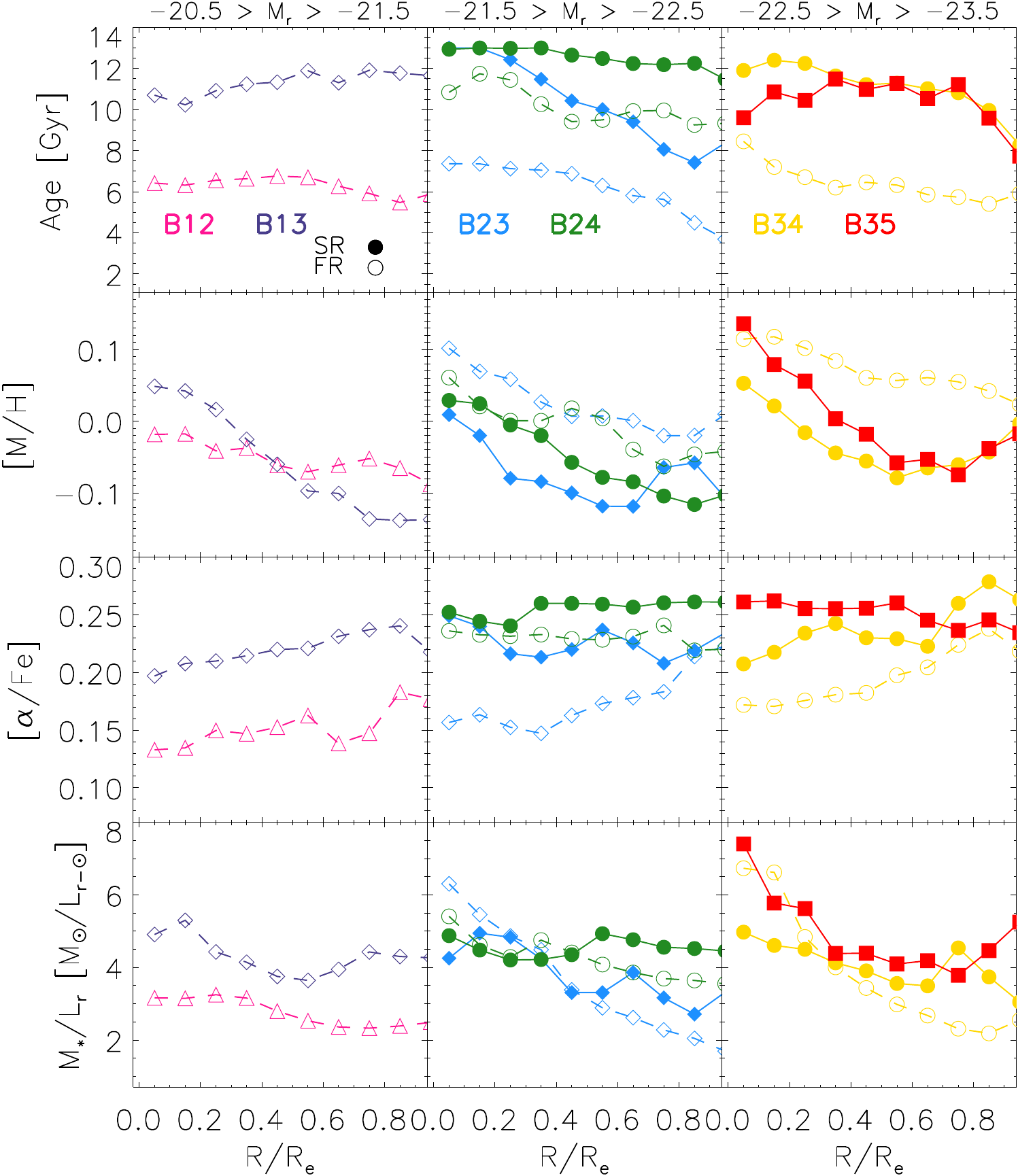}
  \caption{Age, metallicity, $\alpha$-enhancement and M$_*$/L$_r$  versus  $R/R_e$ for the E-SRs and E-FRs (filled and open symbols) when allowing for IMF variations. Colors represent the different bins as stated in the legend, and symbol shape represents $\sigma_0$.  
  } \label{fig:SRFR}
\end{figure}

\subsection{Stellar population gradients}
We now consider the stellar populations.  We estimate SP parameters of the Es exactly as we did for S0s in Section~\ref{sec:grads} and note that although our results allow IMF variations, fixing the IMF to Kroupa results in small quantitative changes, but does not change any of the qualitative trends or conclusions.

Figure~\ref{fig:SRFR} shows the results:  E-SRs (filled symbols) tend to be older, more metal poor (less metal rich) and more $\alpha$-enhanced than E-FRs (open symbols) of the same $M_r$ and $\sigma_0$.  In addition, the left and middle panels show that at fixed $M_r$, E-FRs with large $\sigma_0$ are older, more metal rich (at least in the central regions) and more $\alpha$-enhanced.  Stated differently, at fixed $\sigma_0$, the fainter E-FRs are older, less metal rich and more $\alpha$-enhanced.  These trends are not so obvious, or simply absent for E-SRs.  However, these SP trends for E-FRs are rather similar to those of S0s.  

\begin{figure}
  \centering
  \includegraphics[width=.98\linewidth]{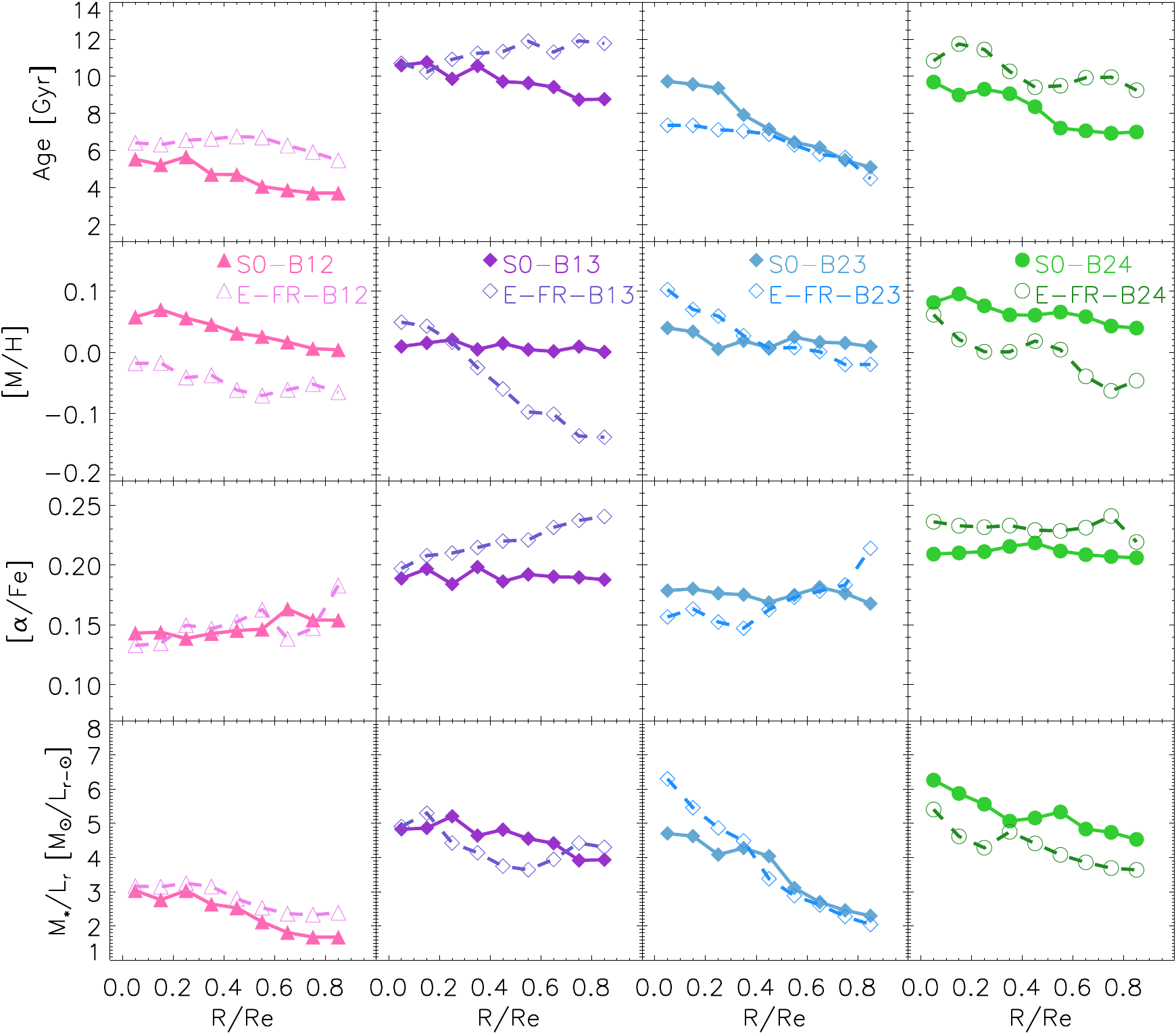}
  \caption{Comparison of the stellar populations of S0s (filled symbols and solid lines) and E-FRs (open symbols and dashed lines) of the same $M_r$ and $\sigma_0$, for a range of values of $M_r$ and $\sigma_0$.  The two panels on the left are fainter than the two on the right, and $\sigma_0$ increases from left to right, except that $\sigma_0$ is the same for the two middle panels. } \label{fig:results-E+S0}
\end{figure}

\begin{figure}
  \centering
  \includegraphics[width=0.9\linewidth]{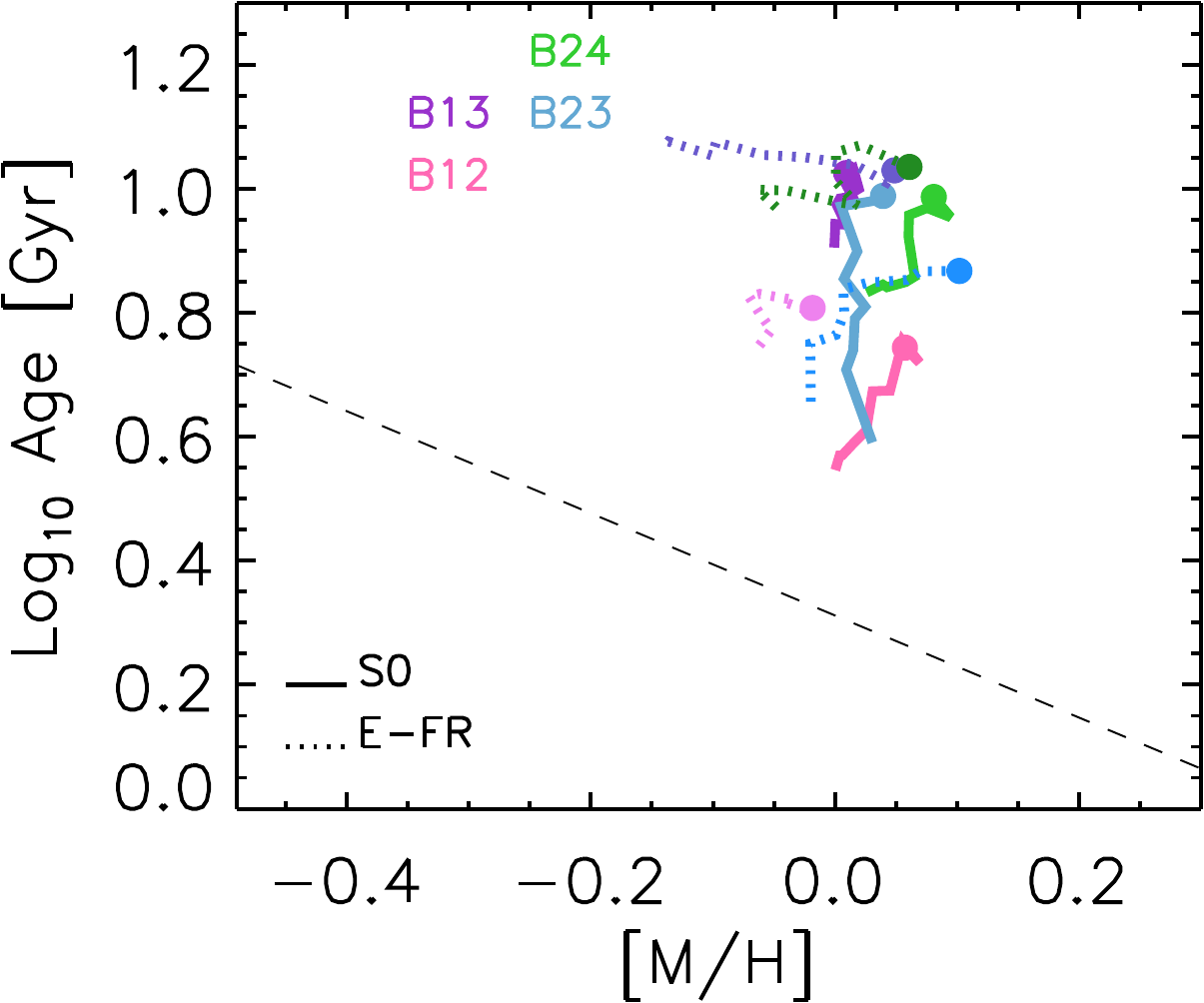}
  \caption{Joint age-metallicity gradients of S0s (solid) and E-FRs (dotted) for the four bins where there is sufficient overlap between the populations (ages and metallicities are same as Figure~\ref{fig:results-E+S0}).  These S0s tend to have stronger age gradients, whereas the corresponding E-FRs tend to have stronger metallicity gradients.} \label{fig:ageMetE+S0}
\end{figure}

\begin{figure*}
  \centering
  \includegraphics[width=0.9\linewidth]{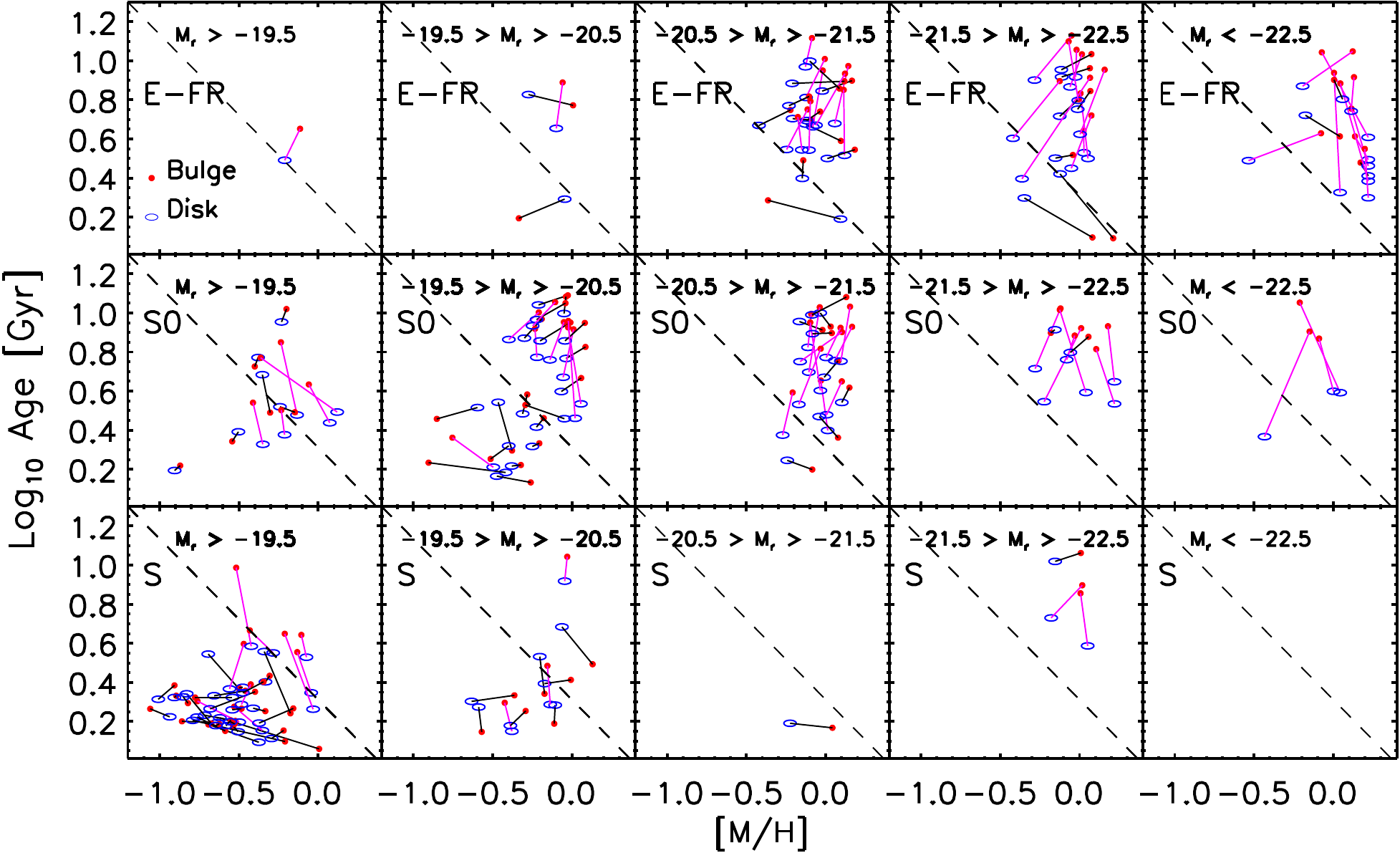}
  \caption{Ages and metallicities for the bulge and disk components of the objects in the sample studied by FM2018, separated according to MDLM-VAC morphology and luminosity:  Bottom to top shows Spirals, S0s and E-FRs, and luminosity increases from left to right.  The three middle luminosity bins are the same as those we focus on in this paper. Magenta lines connect bulges to their disks if the bulge is more than 0.1~dex older, otherwise the line is black. Dashed line, same in all panels, approximately separates the two populations found by FM2018.}\label{fig:misclass}
\end{figure*}

\begin{figure}
  \centering
  \includegraphics[width=0.95\linewidth]{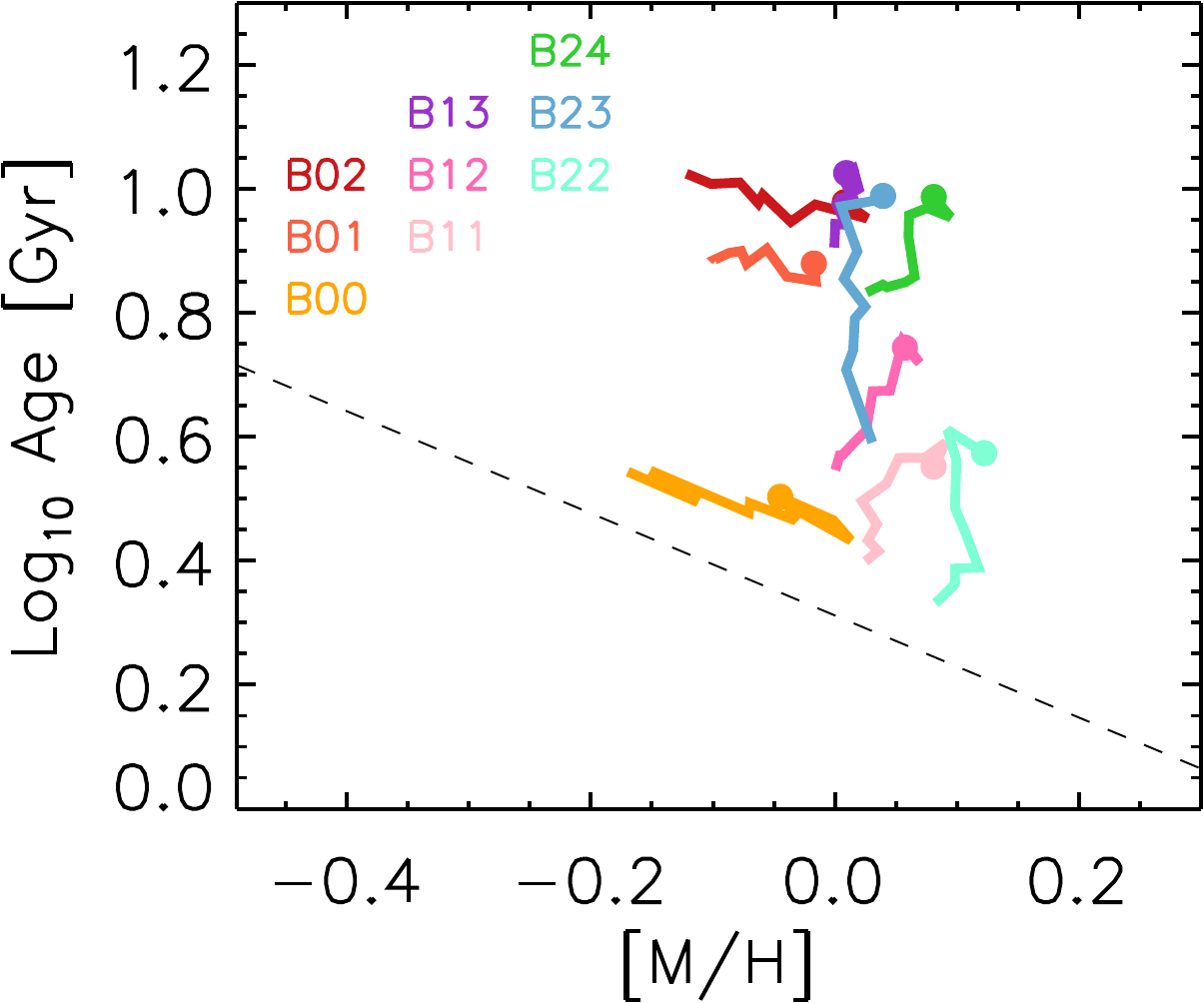}
  \caption{Comparison of age-metallicity gradients of S0s in the nine bins defined in Table~\ref{tab:bin}, color coded as in the previous figures.  Filled circles indicate the central region.   Fainter objects tend to have large metallicity gradients but weak age gradients; luminous objects tend to have large age gradients but weak metallicity gradients.} \label{fig:ageMet}
\end{figure}

Figure~\ref{fig:results-E+S0} presents a direct comparison of the S0s and E-FRs (symbols connected by solid and dashed lines respectively) in bins B12, B13, B23 and B24.  In all but one bin (B23), the S0s tend to be younger and more metal rich.  They also tend to be slightly less $\alpha$-enhanced.  These differences, which are also present if we fix the IMF to Kroupa, are noteworthy, given that they played no role in the morphological classifications.  

Figure~\ref{fig:ageMetE+S0} shows these SP differences in a slightly different format.  The four sets of curves show the joint distribution of age and metallicity from the center (symbols) outwards (tails) for S0s (solid) and E-FRs (dotted) in bins B12, B13, B23 and B24.  This shows that S0s tend to have larger age gradients than their corresponding E-FRs; in contrast, the E-FRs tend to have larger metallicity gradients.  In Paper~II we showed that E-SRs tend to have constant ages (of about 10~Gyrs) with some metallicity gradients, so old E-FRs (bins B13 and B24) are more similar to E-SRs than are S0s.  These systematic SP differences, along with the differences shown in Figure~\ref{fig:sig0sige}, suggest that, contrary to recent claims \citep[e.g.][]{C16, Graham2019}, E-FRs are not just S0s viewed face-on.

If inclination does not drive the differences between them, then what does?  If bulges are dispersion supported and older than disks (which are supported by rotation), then we would expect differences in B/T ratio to be reflected in the kinematic and stellar population gradients.  While E-FRs do have larger B/T (Figure~\ref{fig:sigVrotBTe}), our sample is too small to allow us to test if B/T alone drives the kinematic and SP trends we see.

\section{Comparison with previous work}\label{sec:previous}
Previous work, based on the joint distribution of color, luminosity and velocity dispersion has argued that residuals from the $\sigma_0-L$ relation must be age indicators \citep{Bernardi2005}.  The sense of the age correlation is that at fixed luminosity, older galaxies have larger $\sigma_0$; at fixed $\sigma_0$, older galaxies are fainter \cite[also see Figure~1 of][]{SB09}.  Our Figures~\ref{fig:HbMg} and~\ref{fig:results-rad} show that this is abundantly true for S0s.  Paper~II did not explore a large enough range of $L_r$ to test if this is also true for Es.  Figure~\ref{fig:SRFR} shows that it is, at least for Es that are fast rotators.

The properties of 279 MaNGA S0 galaxies have previously been studied by FM2018.  
They estimate stellar ages and metallicities in bulge and disk regions of individual galaxies, and report a bimodality which strongly correlates with stellar mass: high-mass (M$_*$ $> $10$^{10}$ M$_{\odot}$) S0s are old and metal-rich, with older bulge components; low-mass lenticulars are young and metal-poor without significant differences in the properties of their disk and bulge components.

As we now discuss, our results are in general agreement with theirs, despite significant differences in how we estimate SP parameters.  However, there are a few subtle but important differences which we highlight below.  

The first potentially important difference is in the morphological classification.  FM2018 selected S0 galaxies according to two conditions: galaxies with Galaxy Zoo2 \citep{Willett2013,Hart2016} weighted `smooth' fraction $>$ 0.7 and fast rotating, without any further restriction in luminosity. However, the MDLM-VAC we use here classifies only 45$\%$ of their sample as S0; it classifies 30\% of their sample as Es and 25\% as Spirals \cite[see top panel of Figure~24 in][which shows that Galaxy Zoo2 smooth fraction is not good at discriminating between S0 and other morphologies]{Fischer2019}.  Figure~25 of \citet{Fischer2019} shows that many of the objects with smooth fraction $>0.7$ have Sersic index and bulge fraction values that are consistent with being Spirals, and these are indeed classified as Spirals by the MDLM-VAC. Figure~\ref{fig:badS0s} shows a selection of these objects.

To see how this affects the conclusions of FM2018, Figure~\ref{fig:misclass} shows how the Spirals, S0s and E-FRs (as classified by the MDLM-VAC) in their sample populate the age-metallicity plane, in the format of their Figure~10, for a number of bins in luminosity.  Magenta lines connect bulges to their disks if the bulge is more than 0.1~dex older, otherwise the line is black.  Comparison of the panels shows that the vast majority of the fainter, younger, metal-poor objects which make up FM2018's `second' population are objects classified as Spirals in the MDLM-VAC.  While some of the fainter S0s ($M_r<-20.5$) lie below the dashed line, they are a small fraction of the faint S0s, and so an even smaller fraction of all S0s.  Essentially none of the E-FRs lie below the dashed line.  Evidently, the bimodality reported by FM2018 is primarily due to morphology, rather than luminosity.

Although it is clearly evident in their Figure~10 (and hence in our Figure~\ref{fig:misclass}), FM2018 do not comment on a different sort of bimodality:  Whereas the disks and bulges of massive S0s have essentially the same metallicities (but different ages), the metallicities of disks and bulges can be quite different at lower masses.  We expect this difference to be quite dramatic in our sample, because Figures~\ref{fig:results-sig} and~\ref{fig:results-rad} show that the SP gradients of faint and luminous S0s are very different.  To illustrate, Figure~\ref{fig:ageMet} shows how the SP gradients we measure in our nine bins populate the age-metallicity plane.  Notice the nearly horizontal stripes in the low $L$ bins (centers are less metal poor and very slightly younger -- similar to most faint Spirals in Figure~\ref{fig:misclass}), but vertical stripes at high luminosities (centers are substantially older and only slightly more metal rich, similar to the E-FRs in Figure~\ref{fig:misclass}).  This clear bimodality in S0 properties is obvious even though we are restricting our analysis to the upper end of the mass range covered by FM2018.  (Their bimodality, which Figure~\ref{fig:misclass} suggests is driven by morphology, only really manifests at $M_r>-20$ or fainter.)

\section{Discussion and conclusions}\label{sec:discuss}
We used stacked spectra to study how the kinematics of and chemical abundances in S0 galaxies vary with velocity dispersion and luminosity (Figure~\ref{fig:sample}).   We interpreted the chemical abundances using (MILES bimodal) single stellar population models.  

$\bullet$ Our analysis has uncovered a bimodality in the S0 population.  The correlations between size or velocity dispersion and stellar mass change slope at around $3 \times 10^{10}M_\odot$ (left hand panel of Figure~\ref{fig:VL}).  This coincides with the mass scale at which gradients in kinematics and stellar population change dramatically.  In particular, there are obvious gradients in line-index strength (Figures~\ref{fig:HbMg}--\ref{fig:TiO2}) which also change at this mass scale.  When fit to our measurements, single stellar population models indicate that the more massive S0s show strong age gradients but little or no metallicity gradient, while the less massive ones present relatively flat age and velocity dispersion profiles, but a significant metallicity gradient (Figure~\ref{fig:ageMet}).

$\bullet$ S0s with the smallest $\sigma_0$ in each luminosity bin -- i.e., the youngest S0s for their $L_r$ -- have IMFs that are closest to Kroupa (Figure~\ref{fig:IMF}) -- but IMFs are otherwise a little more bottom heavy.  (We accounted self-consistently for IMF differences when estimating other SP parameters, which are all light- not mass-weighted.)

$\bullet$ While at fixed $L_r$, age increases with $\sigma_0$ and metallicity decreases as expected, we find that, at fixed $\sigma_0$, more luminous S0s are younger, more metal rich and less $\alpha$-enhanced (compare same symbols in the different panels of Figure~\ref{fig:results-rad}). Evidently, the usual statement that `massive galaxies are older' is not true if $\sigma_0$ is held fixed.
  
$\bullet$ We also found differences between the kinematic profiles and stellar populations of the S0 and E-FR populations.  The velocity dispersion profiles of (face-on) S0s are steeper than those of E-FRs (Figure~\ref{fig:sig0sige}), despite the fact that both populations have similar rotation (top right panel of Figure~\ref{fig:sigVrotBTe}).  This indicates that S0s tend to be less supported by dispersion than are E-FRs.  Like S0s -- but in contrast to E-SRs -- at fixed $\sigma_0$, more luminous E-FRs are younger and more metal rich (Figure~\ref{fig:SRFR}).  However, their gradients differ:  S0s tend to have larger age gradients and smaller metallicity gradients than E-FRs of the same $L$ and $\sigma_0$ (Figure~\ref{fig:ageMetE+S0}).  The age-metallicity gradients of E-FRs are more similar to those of E-SRs.  Together, these SP and kinematic differences suggest that distinguishing between E-FRs and S0s is meaningful:  E-FRs are not just S0s viewed face-on. While E-FRs tend to have larger B/T ratios (bottom left panel of Figure~\ref{fig:sigVrotBTe}), our sample is not large enough to determine if the differences between E-FRs and S0s are entirely driven by B/T.

Gradient strength is thought to be an indicator of the interplay between star formation and stellar mass assembly.  The qualitative change in gradients that we see in the S0 population, coupled with the change in global scaling relations at $M_*\sim 3\times 10^{10}M_\odot$, suggest that this is a special scale for S0s.  It is tempting to assert that, below this mass scale, S0s are closer to Spirals -- mergers are less important than in situ star formation and perhaps, some gentle gas stripping.  These galaxies have metallicity gradients but little age gradient. They also have similar metallicities but with the larger $\sigma$ S0s being older and more $\alpha$-enhanced.  However, the transition from metallicity to age gradients just above this mass scale suggests gas rich mergers are beginning to matter, as these would result in more recent star formation from enriched gas, and hence younger light-weighted ages with higher metallicity and smaller $\alpha$-enhancement -- this would explain why at fixed $\sigma_0$, more luminous S0s and E-FRs are younger, more metal rich and less $\alpha$-enhanced (compare same symbols in the different panels of Figure~\ref{fig:results-rad}). In addition, if the merger remnant is able to grow a disc, it will have higher metallicity, younger age and lower [$\alpha$/Fe] in its outer regions compared to its progenitors, thus explaining the change in S0 gradients we see as we cross $M_*\sim 3\times 10^{10}M_\odot$ (Figure~\ref{fig:ageMet}).

Mergers increase the stellar mass, but the abundance of S0s drops sharply above 
$2\times 10^{11}M_\odot$ (Figure~\ref{fig:VL}); this mass scale was first identified as being important by \citet{Bernardi2011}.  More recent work \cite[][and our Papers~I and~II]{Cappellari2013b} has confirmed its significance.  At these higher masses there is a mix of E-FRs and E-SRs.  E-FRs may result from mergers which do not produce discs, or for which a disc would have been unstable. Younger E-FRs (pink and blue dotted lines in Figure~\ref{fig:ageMetE+S0}) have age-metallicity gradients which are intermediate between those of S0s and older Es (Figure~\ref{fig:SRFR}). Older E-FRs (purple and green dotted lines in Figure~\ref{fig:ageMetE+S0}) have metallicity decreasing outwards from the center, but little if any age gradient within $R_e$.  This is similar to E-SRs -- which are uniformly old, have metallicity gradients and are thought to be dominated by dry merger assembly histories.  However, these E-FRs tend to be younger, more metal rich and less $\alpha$-enhanced than E-SRs of the same $L$ and $\sigma_0$ (Figure~\ref{fig:SRFR} and Paper~II), consistent with E-FRs having had more extended star formation histories.

\section*{Acknowledgements}
We are grateful to the referee for a helpful report, and to A. Fraser McKelvie for sharing the data used in FM2018. This work was supported in part by NSF grant AST-1816330. HDS acknowledges support from Centro Superior de Investigaciones Científicas PIE2018-50E099. FN acknowledges support from the National Science Foundation Graduate Research Fellowship (NSF GRFP) under Grant No. DGE-1845298. 

Funding for the Sloan Digital Sky Survey IV has been provided by the Alfred P. Sloan Foundation, the U.S. Department of Energy Office of Science, and the Participating Institutions. SDSS acknowledges support and resources from the Center for High-Performance Computing at the University of Utah. The SDSS web site is www.sdss.org.

SDSS is managed by the Astrophysical Research Consortium for the Participating Institutions of the SDSS Collaboration including the Brazilian Participation Group, the Carnegie Institution for Science, Carnegie Mellon University, the Chilean Participation Group, the French Participation Group, Harvard-Smithsonian Center for Astrophysics, Instituto de Astrof{\'i}sica de Canarias, The Johns Hopkins University, Kavli Institute for the Physics and Mathematics of the Universe (IPMU) / University of Tokyo, Lawrence Berkeley National Laboratory, Leibniz Institut f{\"u}r Astrophysik Potsdam (AIP), Max-Planck-Institut f{\"u}r Astronomie (MPIA Heidelberg), Max-Planck-Institut f{\"u}r Astrophysik (MPA Garching), Max-Planck-Institut f{\"u}r Extraterrestrische Physik (MPE), National Astronomical Observatories of China, New Mexico State University, New York University, University of Notre Dame, Observat{\'o}rio Nacional / MCTI, The Ohio State University, Pennsylvania State University, Shanghai Astronomical Observatory, United Kingdom Participation Group, Universidad Nacional Aut{\'o}noma de M{\'e}xico, University of Arizona, University of Colorado Boulder, University of Oxford, University of Portsmouth, University of Utah, University of Virginia, University of Washington, University of Wisconsin, Vanderbilt University, and Yale University.

\renewcommand\thefigure{A\arabic{figure}}
\setcounter{figure}{0}

\begin{figure*}
  \centering
  \includegraphics[width=0.8\linewidth]{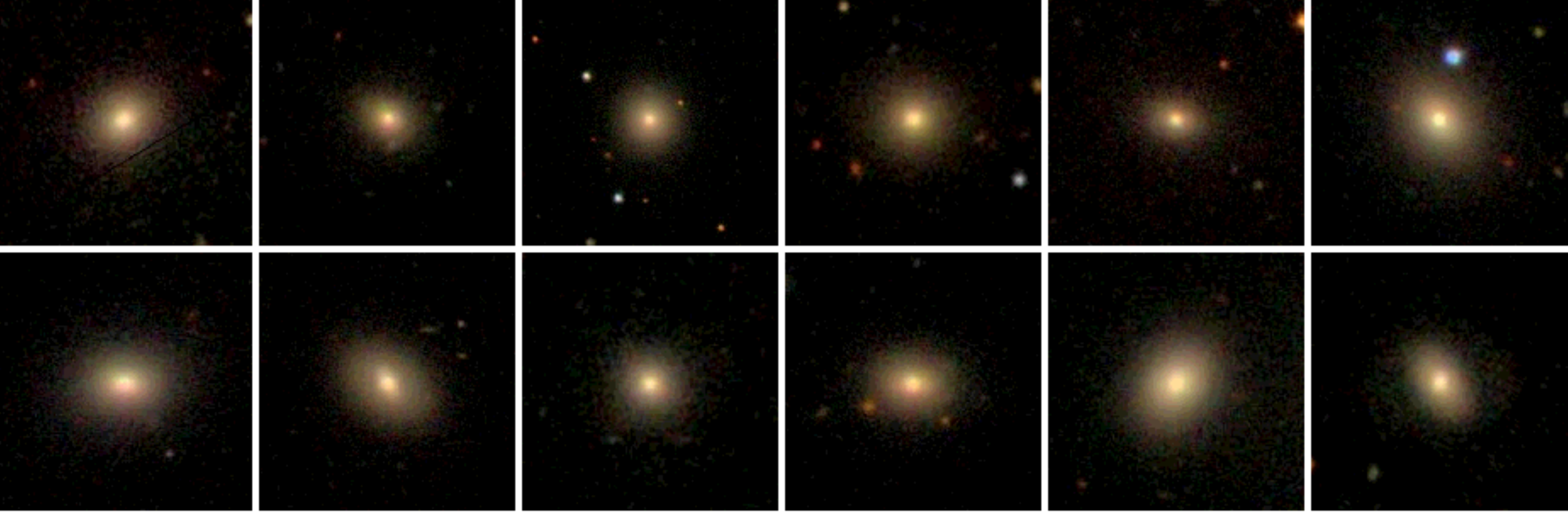}
  \caption{Random selection of objects which we classify as E-SRs in the two luminosity bins in common with the S0 sample: $-20.5 > M_r > -21.5$ (top) and $-19.5 > M_r > -20.5$ (bottom).}
  \label{fig:E-SRs}
  \includegraphics[width=0.8\linewidth]{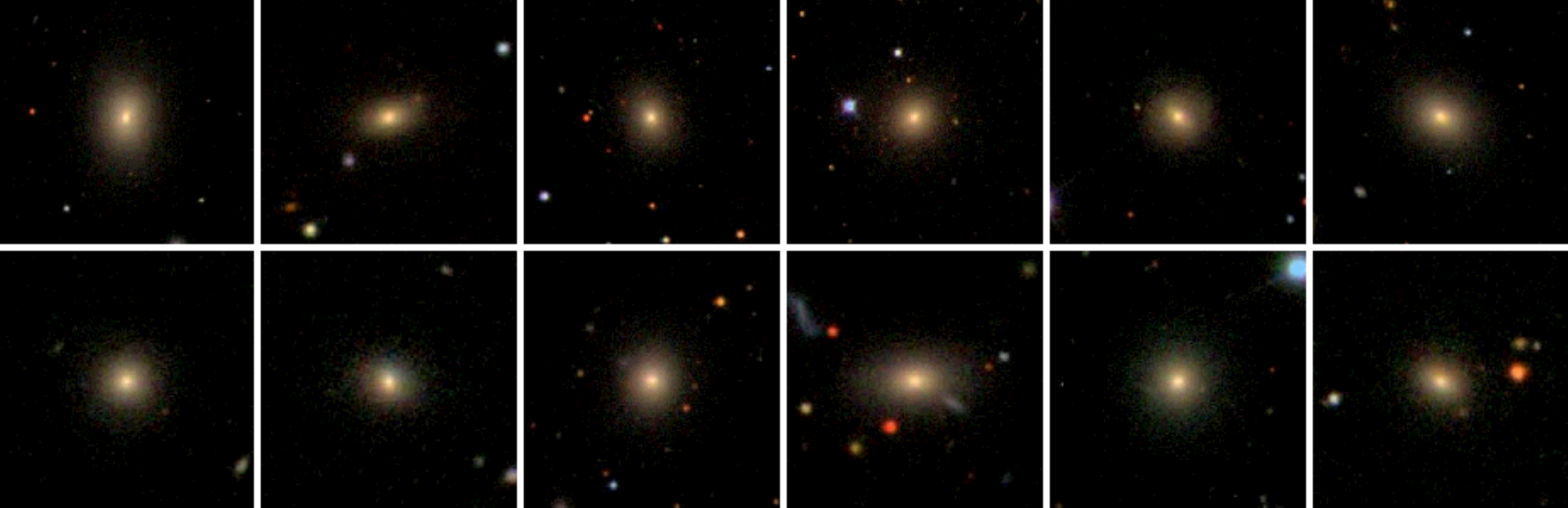}
  \caption{Same as previous figure, but for E-FRs.}
  \label{fig:E-FRs}
   \includegraphics[width=0.8\linewidth]{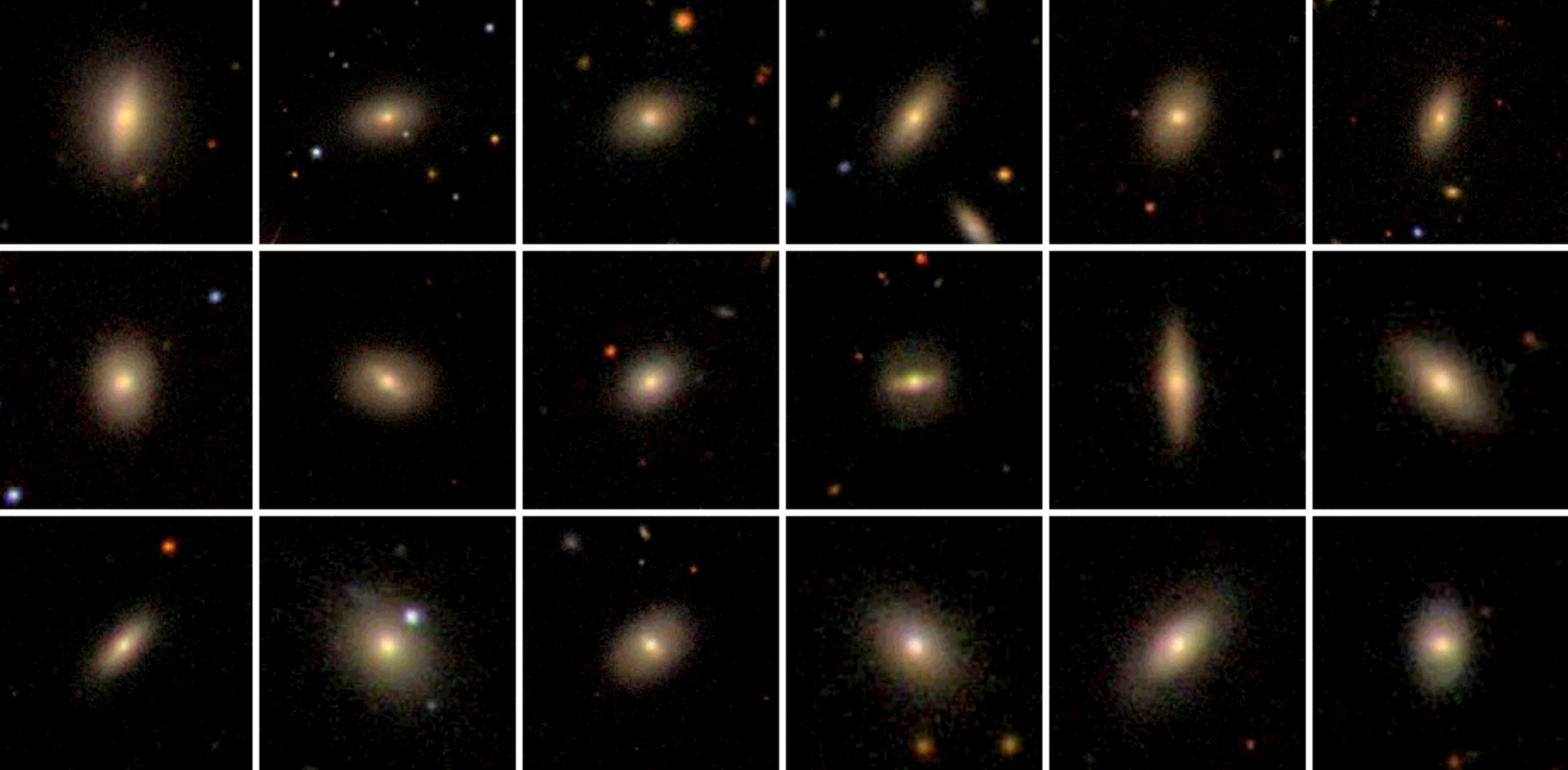}
  \caption{Random selection of objects which we classify as S0s in the three luminosity bins used for our analysis: $-20.5 > M_r > -21.5$ (top) and $-19.5 > M_r > -20.5$ (middle) and $-19.5 > M_r > -20.5$ (bottom). }
  \label{fig:S0s}
\end{figure*}

\begin{figure*}
  \centering
  \includegraphics[width=0.9\linewidth]{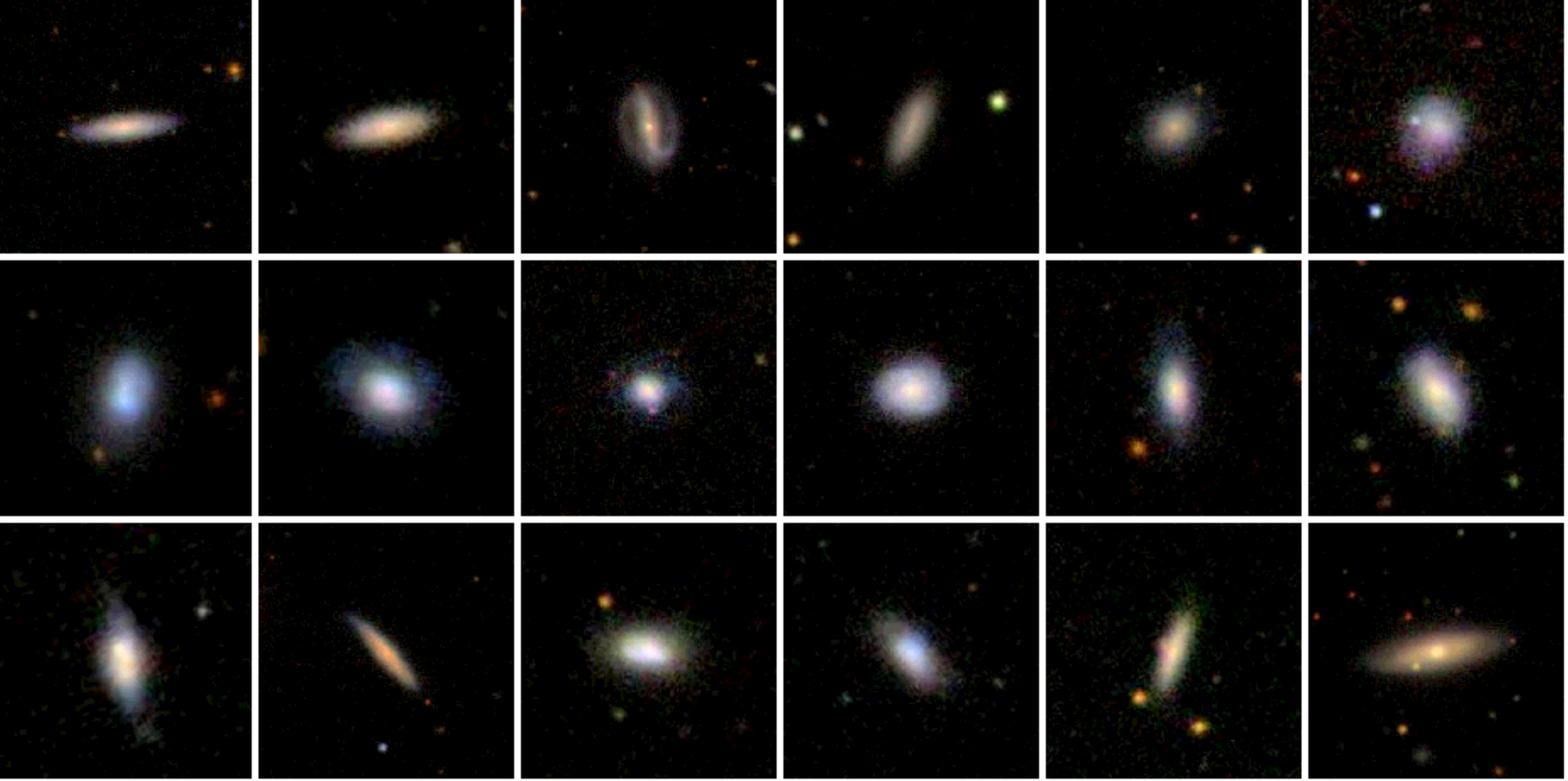}
  \caption{Random selection of objects which FM2018 classify as S0s but we classify as Spirals. A considerable fraction of these objects are indeed Spirals.  }
  \label{fig:badS0s}
\end{figure*}





\bibliographystyle{mnras}
\bibliography{biblio} 




\appendix
\section{Images}
Morphology plays an important role in our sample selection.  
This Appendix shows a random selection of objects to illustrate that our morphological classifications are reasonable.  Comparison of Figures~\ref{fig:E-SRs} and~\ref{fig:E-FRs} shows that E-SRs and E-FRs are very similar morphologically, whereas comparison of Figures~\ref{fig:E-FRs} and~\ref{fig:S0s} shows that S0s tend to have more obvious disks.  

Finally, Figure~\ref{fig:badS0s} shows a random selection of objects which FM2018 classify as S0s but we classify as Spirals.  A considerable fraction of these objects are indeed Spirals, suggesting that the FM2018 S0 sample is rather impure.  As not all of these objects are obviously Spirals, our S0 sample may be slightly incomplete.  However, we believe it is much more pure.  


\bsp	
\label{lastpage}
\end{document}